\documentstyle[aas2pp4]{article}
\def\simlt{\hbox{ \rlap{\raise 0.425ex\hbox{$<$}}\lower 0.65ex\hbox{$\sim$} }}
\def\simgt{\hbox{ \rlap{\raise 0.425ex\hbox{$>$}}\lower 0.65ex\hbox{$\sim$} }}

\begin{document}

\lefthead{Makino and Hut}
\righthead{Merger Rate}

\title{Merger Rate of Equal-Mass Spherical Galaxies}

\author{Junichiro Makino}

\affil{Department of Information Science and Graphics,\\
College of Arts and Sciences, University of Tokyo,\\
3-8-1 Komaba, Meguro-ku, Tokyo 153, Japan.}

\and

\author{Piet Hut}
\affil{Institute for Advanced Study, Princeton, NJ 08540}

\def\versiondate{{7 Nov. 1996}}
\def\titleabbrev{{Merger Rate}}
%
%
\def\undertext#1{{$\underline{\hbox{#1}}$}}
\def\doubleundertext#1{{$\underline{\underline{\hbox{#1}}}$}}
\def\half{{\scriptstyle {1 \over 2}}}
\def\ie{{\it {\frenchspacing i.{\thinspace}e. }}}
\def\eg{{\frenchspacing e.{\thinspace}g. }}
\def\cf{{\frenchspacing\it cf. }}
\def\etal{{\frenchspacing\it et al.}}
\def\et{{\etal}}
\def\simlt{\hbox{ \rlap{\raise 0.425ex\hbox{$<$}}\lower 0.65ex\hbox{$\sim$} }}
\def\simgt{\hbox{ \rlap{\raise 0.425ex\hbox{$>$}}\lower 0.65ex\hbox{$\sim$} }}
\def\solar{\odot}
\def\msun{\ifmmode{M_\solar}\else{$M_\solar$}\fi}
\def\rsun{\ifmmode{R_\solar}\else{$R_\solar$}\fi}
\def\Rf{\parindent=0pt\smallskip\hangindent=3pc\hangafter=1}
\def\pc{{\rm pc}}
\def\kpc{{\rm kpc}}
\def\Mpc{{\rm Mpc}}
\def\yr{{\rm yr}}
\def\Myr{{\rm Myr}}
\def\Gyr{{\rm Gyr}}
\def\kT{\ifmmode{kT}\else{$kT$}\fi}
\def\N{{\ifmmode{N}\else{$N$}\fi}}
\def\fb{{\ifmmode{f_B}\else{$f_B$}\fi}}
\def\emax{{\ifmmode{E_{max}}\else{$E_{max}$}\fi}}
\def\td{{\ifmmode{t_d}\else{$t_d$}\fi}}
\def\tcr{{\ifmmode{t_{cr}}\else{$t_{cr}$}\fi}}
\def\tr{{\ifmmode{t_r}\else{$t_r$}\fi}}
\def\trh{\ifmmode{t_{rh}}\else{$t_{rh}$}\fi}
\def\vv{{\ifmmode{\langle v^2\rangle}\else{$\langle v^2 \rangle$}\fi}}
\def\v{{\ifmmode{\langle v^2\rangle^{1/2}}
		\else{$\langle v^2 \rangle^{1/2}$}\fi}}
\def\half{{\ifmmode{{1 \over 2}}\else{${1 \over 2}$}\fi}}
\def\dhalf{{\textstyle {1 \over 2}}}
\def\threehalf{{\ifmmode{{3 \over 2}}\else{${3 \over 2}$}\fi}}
\def\dthreehalf{{\textstyle {3 \over 2}}}
\def\dfivehalf{{\textstyle {5 \over 2}}}
\def\dfivethree{{\textstyle {5 \over 3}}}
\def\kms{\ifmmode{\rm km\,s^{-1}}\else{$\rm km\,s^{-1}$}\fi}
\def\kmps{{\rm km/s}}
\def\piet#1{{\bf[#1 -- piet]}}
\def\jun#1{{\bf[#1 -- jun]}}
%
%
\font\lgh=cmbx10 scaled \magstep2
\baselineskip 15pt
\parskip 0pt plus 2pt
%
%
%
%

\begin{abstract}

We present cross sections and reaction rates for merging to occur
during encounters of equal-mass spherical galaxies.  As an
application, we determine the rate of galaxy merging in clusters of
galaxies.  We present results for two types of Plummer models (a full
and a truncated one), two King models and the Hernquist model.  Cross
sections are determined on the basis of a large number ($\sim 500$) of
simulations of galaxy encounters, using the 10-Gigaflops GRAPE 3A
special-purpose computer.  We characterize the overall merger rate of
galaxies in a galaxy cluster by a single number, derived from our
cross sections by an integration over galaxy encounter velocities in
the limit of a constant density in velocity space.  For small
clusters, where the cluster velocity dispersion may not significantly
exceed the internal velocity dispersion of the individual galaxies,
this constant-density approximation may not be valid.  For those
cases, we present separate results, based on integrations of our cross
sections over Maxwellian velocity distributions.  Finally, tidal
effects from the cluster potential as well as from neighboring
galaxies may prevent a barely bound galaxy pair from spiraling in
after their first encounter.  We give a quantitative estimate of the
resulting reduction in the actual merger rate, due to these tidal
interactions.

\end{abstract}

\newpage

\section{Introduction}

Detailed simulations of galaxy encounters have become increasingly
sophisticated during the last twenty-five years, due to significant
improvement both in computer hardware and in algorithms used (for a
recent review, see \cite{Barnes1992}).  Most of these
simulations attempt to explain features of specific encounters, often
in an attempt to reproduce particular observations of interacting
galaxies or merger remnants.  In contrast, encounters between simpler
galaxy models in order to obtain more general statistical results,
have received less attention.

The present paper attempts to address this latter problem, by limiting
ourselves to simulations of equal-mass galaxy models of various types,
in order to determine merger cross sections and reaction rates.  Our
results can be readily applied to study the evolution of galaxy
clusters, if we use our models to represent dark matter halos around
galaxies in the limit that these halos can be considered to be
spherical.  In practice, moderate deviations from halo sphericity will
not greatly affect our results.

The merger probability in an encounter of two galaxies is enhanced
significantly when the encounter takes place at relatively low speed.
For a given cluster of galaxies, with a given population of specified
galaxy halos, the net merger rate can be determined by an integration
of the usual $n\sigma v$ factor (density $\times$ cross section
$\times$ velocity) over a Maxwellian or lowered Maxwellian velocity
distribution.  This procedure is sketched qualitatively in Fig. 1b,
and can be compared to that used in stellar evolution calculations
(Fig. 1a) where nuclear reaction rates are determined from the
encounters of individual nuclei in the cores of stars.  In the latter
case, cross sections rise quickly with increasing encounter speed,
leaving a small window for effective encounters well into the
high-energy tail of the Maxwellian distribution.  Our case of galaxy
mergers, in contrast, is dependent on low-speed encounters, and
determined by a relatively small window at the low-energy end of the
Maxwellian distribution.

Fig. 1b suggests that we can make a good start by representing the
velocity distribution of the galaxies in the cluster as having a
constant density in velocity space, \ie\ $f(v) \propto v^2$ in three
dimensions.  This constant density can be obtained, for any given
Maxwellian distribution, as the density of galaxies around the peak of
the folded merger occurrence $f\sigma v$ in Fig. 1b.  In \S 2 we
describe some technical details of our calculations.  In \S 3 we
present the cross sections, from which we determine reaction rates in
\S 4.  In \S 5, we determine the corrections to the reaction rates,
due to tidal effects.  As an application, we present the rate of galaxy
merging in clusters of galaxies in \S 6, and compare that with results
from previous papers.  \S7 sums up.

\section{Simulations}

\subsection{Units}

We choose our physical units of mass, length, and time by requiring
that $M=r_v=G=1$, where $M$ is the mass on a single galaxy, $r_v$ is
the virial radius of a galaxy and $G$ is the gravitational constant
(\cite{Heggie1986}).  The virial radius is a measure of the size
of a galaxy, defined so as to be able to express the potential energy
of the galaxy to be equal to
\begin{equation}
E_{pot} = - {G M^2 \over 2 r_v}
\end{equation}
In practice, the virial radius is not very different from the
half-mass radius $r_h$.  Typically, 
\begin{equation}
r_h \simeq 0.8 r_v,
\label{somethingwith0.8}
\end{equation}
with a
coefficient generally in the range 0.76 -- 0.98 (\cite{Spitzer1987}, \S
1.2a).  In our standard case of a Plummer model we have
\begin{equation}
r_h = {3\pi \over 16  \sqrt{2^{2/3} - 1}} r_v = 0.77r_v,
\end{equation}
for the Hernquist model
\begin{equation}\label{eqn2:4}
r_h = {(1 + \sqrt2) \over 3}r_v,
\end{equation}
and for a constant-density sphere
\begin{equation}
r_h = { 3\times 2^{2/3}\over 5} r_v = 0.95 r_v.
\end{equation}

The main reason to take the virial radius, rather than the more usual
half-mass radius, as our choice of units is a physical one: when we
vary the internal structure of our galaxies, it is more meaningful to
keep the total energy $E$ of a galaxy constant than its half-mass
radius.  In our units,
\begin{equation}
E = E_{pot}/2 = -1/4.
\label{somethingwith0.25}
\end{equation}

In terms of the particle positions, the virial radius is the harmonic
mean particle separation
\begin{equation}
r_v = \left< {1 \over r_i - r_j} \right>^{-1},
\end{equation}
averaged over all particle pairs ${i \not= j}$.

\subsection{Galaxy models and initial conditions}

For the initial spherical galaxy model, we adopted five different
choices: two Plummer models with different cutoff radii $r_c$ (22.8
and 4), two King models with different values for the central
potential $W_c$ (1 and 7), and a Hernquist model (\cite{Hernquist1990}) with
$r_c = 20$.  We will indicate the Plummer model with $r_c = 4$ by
`truncated Plummer' and the Plummer model with $r_c = 22.8$ simply by
`Plummer'.

These models provide a wide range of central condensation and halo
distribution.  They all have an isotropic velocity distribution since,
according to Aguilar and White (1985), the difference in the velocity
distribution is less important in determining merger cross sections
than the difference in the density distribution. For all calculations,
initial galaxies are modeled by 2048 equal-mass particles (except for
several test calculations, see \S 3.4). Thus the total number of
particles per run is 4096. For Plummer models and King models, we
constructed random realizations using an algorithm similar to the one
described in Aarseth, Henon and Wielen (1974). For Hernquist models,
we adopted a kind of ``quiet start'', in which particle $i$ is
initially placed in an arbitrary position on a sphere with a radius
corresponding to the Lagrangian mass $(i-0.5)/N$.  The velocity of
each particle was then chosen randomly from the velocity distribution
function at that radius.

Since the initial galaxy models are spherical, the relative orbit of
the two given galaxies is determined by only two
parameters specifying the incoming branch of the hyperbolic Kepler
orbit, the impact parameter $\rho$ and the relative velocity at infinity
$v$.  In this study, we found it more convenient to specify the
initial orbit by $v$ and the pericenter distance $r_p$ of the
extrapolated unperturbed hyperbolic orbit, in the limit that two
galaxies would have been point masses. The reason to use $r_p$ rather
than $\rho$ is related to the measurements of energy change during
relatively wide encounters: keeping $r_p$ constant allows us to let
$v$ smoothly go to zero in the parabolic limit, where $\rho
\rightarrow \infty$.  

Fig. 2 shows the initial conditions used to determine the merging
criterion for each models. The number of runs for one model is 60-140.
The following values for $r_p$ were used: 0, 0.0125, 0.05, 0.2, 0.4,
0.6, 0.8, 1.2, 1.4, 1.6, 1.8, 2.0, 2.4, 2.8, 3.2, 3.6, 4.0, 4.8, 5.6,
6.4, 8.0.  For the initial separation between the two approaching
galaxies, we choose a value of roughly $10\max(r_{p}, 2r_h)$, where
$r_h$ is the half mass radius of each of the two identical galaxies.

\subsection{Hardware and Numerical Integration Method}

We have used a GRAPE-3A (\cite{Okumura1993}), a special-purpose
computer for collisionless $N$-body simulation, for all simulations.
The GRAPE-3A performs the calculation of the gravitational interaction
between particles.  It is used with a host computer which is a
UNIX-based workstation.  All calculations except for the force
calculation are performed on the host computer.  One GRAPE-3A board
hosts eight custom LSI chips, each of which calculates the
gravitational interaction between particles in the speed of about 0.75
Gflops. Thus the peak speed of one board is 6 Gflops. For most runs,
we have used two GRAPE-3A boards in parallel.

For the force calculation, we have used a simple direct-summation
algorithm.  Although it is perfectly possible to use the $O(N \log N)$
tree algorithm on the GRAPE systems (\cite{Makino1991}), it is not practical
to do so for small particle numbers.  The minimum number of particles
for which the tree code is faster than the direct summation algorithm
is $1\sim 4 \times 10^4$, depending on the speed of the host computer.
Since we typically used only 4096 particles, direct summation was much
faster than the tree code.

We used the time-centered leap-frog integrator.  The time step was
1/128 for all runs.  The size of the softening parameter was $\epsilon
= 1/32$.  These parameters gave sufficient accuracy, resulting in a
relative energy error well below 0.1 \% at the end of each
calculation).

\subsection{Simulation Termination Procedure and Data Reduction}

For each choice of galaxy model, and for each choice of pericenter
distance $r_p$, we have carried out a series of galaxy encounter
simulations at different values for the encounter velocity $v$
(measured at infinity).  For each of these simulations, we determine
the energy of the relative orbit of the two galaxies, well after the
first encounter, using the simple energy criterion:
\begin{equation}\label{eqn2:7}
E = -{GM_1M_2 \over r} + {1 \over 2}\mu v^2,
\end{equation}
where $M_1$ and $M_2$ are the masses of two galaxies, $r$ is the
distance between two galaxies, $G$ is the gravitational constant,
$\mu=M_1M_2/(M_1+M_2)$ is the reduced mass, $v$ is the relative velocity.

In Fig. 3, we plot the energy $E(t)$ as a function of time $t$, for
the case of Hernquist model initial conditions, with $r_p = 0.0125$
and $v=1.2$.  It is clear that the asymptotic value of the
energy is reached soon after the encounter.  However, to be on the
safe side, we have extended all runs to reach a final separation
comparable to the initial separation, where possible (for a bound final
state, we have taken the minimum of this separation and the apocenter
of the elliptic orbit).  In this particular example, it takes far more
time to reach the initial separation again, because the outgoing orbit
has far less positive energy and therefore the galaxies recede more
slowly than they came in.

Having determined the asymptotic energy values $E(r_p, v)$, we
select the pair of $v$ values between which the energy changes
sign.  We then use linear interpolation to determine the critical
point $r_p(v)$ at which the outgoing orbit would have been just
parabolic, separating the regions of merging and escape.  Fig. 2
indicates this process, for the case of Plummer models.  Each circle
represents a galaxy encounter run, and the full line connects the
critical $r_p(v)$ points.

Fig. 4 shows the same results, but translated from $r_p$ values to
$\rho$ values.  The impact parameter $\rho$ is related to the
pericenter distance $r_p$ through the gravitational focusing relation
(see Eq. \ref{eqn3:5} below).

As a technical note, here is a brief description of the actual
implementation of the energy determination, Eq. (\ref{eqn2:7}), in which
appear the mass, position and velocity of the galaxies after the
encounter. They are calculated by the following procedure:

\begin{description}

\item{1)} Make a crude guess for the center of mass of each galaxies.
Here, we use the center of mass of particles that are initially in one
galaxy as the guess for the center of mass of the galaxy at that time.

\item{2)} For each particle, determine which galaxy it belongs to. We
calculate estimated binding energy given by
\begin{equation}
E_{ij} = -{GM_j \over (r_{ij}^2 + R_0^2)^{1/2}}  + 
{1 \over 2}v_{ij}^2,
\end{equation}
where $r_{ij}$ is the distance between particle $i$ and the estimated
center of mass of the galaxy $j$, $v_{ij}$ is the relative velocity
between them, and $R_0$ is the length scale parameter that represents
the depth of the potential well of galaxies. We use $R_0 = 0.6$, which
gives a fairly accurate estimate for the potential for Plummer models.
We then make our initial guess as to which galaxy particle $i$ mostly
likely might be bound, by assigning it to galaxy $1$ if  $E_{i1}
< E_{i2}$, or to galaxy $2$ if  $E_{i1} > E_{i2}$.

\item{3)} For each galaxy, determine which of the particles that are
labeled as belonging to it are actually bound to the galaxy. If
the binding energy of a particle relative to the galaxy is positive,
we regard it as an unbounded escaper.

\item{4)} Repeat step (3) until membership converges.

\end{description}

Most of the data reduction was performed using the NEMO software
package, in the version provided by Peter Teuben (1995).

\section{Cross Sections for Galaxy Merging}

\subsection{Results and Scaling Relations}

Fig. 5 shows the critical velocity for merging $v_{crit}$ as a
function of the impact parameter $\rho$, for all galaxy models used.
The merging criterion determined experimentally here shows only a weak
dependence on galaxy model: for the whole range of impact parameter
studied, the difference in $v_{crit}$ among the different models is
less than 20\%.  For the Plummer model, we find a good fit with
$v_{crit}\propto \rho^{-0.75}$ for large $\rho$, as plotted for
comparison in Fig. 5.  For the King model with $W_c=1$ the index is
slightly larger and for $W_c=7$ the index is slightly smaller.  The
slope for the Hernquist model is somewhat more small.

At first sight, these results seem to be in good agreement with the
distant tidal impulsive approximation, which indeed predicts a
relationship $v_{crit}\propto \rho^{-0.75}$ for large $\rho$.  The
tidal impulsive approximation has been used by many researchers to
obtain analytic approximations for changes in mass and energy of
galaxies involved in relatively weak encounters (\cite{Richstone1975,White1979,Dekel1980,Richstone1983}).  In
this approximation, the loss of energy of the relative orbital motion
of the two galaxies scales as
\begin{equation}
\Delta E \propto \left({GM \over r_p^2v_p}\right)^2,
\end{equation}
where $r_p$ and $v_p$ is the relative distance and relative velocity
at closest approach. For the critical velocity, the energy loss is
equal to the kinetic energy of the orbital motion at the infinity:
\begin{equation}
\Delta E = {1 \over 2} \mu v^2,
\end{equation}
where $\mu = M/2$ is the reduced mass. If we apply the point mass
approximation for the relative orbit, the relations between the orbital
elements are given by conservation of angular momentum and energy as
\begin{eqnarray}
\rho v &=& r_pv_p,\nonumber\\
{1 \over 2} \mu v^2 &=& {1 \over 2} \mu v_{p}^2 - {GM^2 \over
r_p},
\end{eqnarray}
Here, we assume that $\mu v^2 << GM^2/r_p$. Under this
assumption, the last equality is reduced to
\begin{equation}
v_p = \sqrt{4GM \over r_p}.
\end{equation}
The resulting pericenter distance is then given as
\begin{equation}\label{eqn3:5}
r_p = {\rho^2v^2 \over 4GM}.
\end{equation}
and the incoming velocity can now be expressed in terms of the impact
parameter as
\begin{equation}
v \propto \rho^{-3/4},
\end{equation}
seemingly in good agreement with the data shown in Figure 6.

\subsection{The Masking Tendency of Gravitational Focusing}

This good agreement is, however, fortuitous and does not reflect the
physical reality.  The discrepancies can be seen clearly in Fig. 6,
which shows the critical velocity $v_m$ as a function of pericenter
distance $r_p$.  If the tidal approximation were really valid, the
results for all models would show the same asymptotic behavior of \ $v
\propto r_p^{-3/2}$. The results for the Plummer models and the
concentrated King model is close to the theoretical line. However, the
results for the Hernquist model and the shallow King model deviate
significantly.

Indeed, there was no reason to expect an impulsive tidal approximation
to give us guidance in estimating the critical velocity for mergers
to take place.  As is clear from Fig. 6, mergers at large $r_p$ take
place for nearly parabolic orbits, where the duration of relatively
close encounters around pericenter passage is drawn out to a total
time exceeding that of the half-mass crossing time.  There is no
reason to expect, under these circumstances, that any type of
impulsive approximation would be valid.

Why, then, did Fig. 5 show such a good correspondence with the line
$v_{crit}\propto \rho^{-0.75}$?  The reason can be found in a
conspiracy of gravitational focusing effects together with the real
merger criterion.  This can be seen as follows.

Suppose that we start with a relation for the energy loss in the form
of a power law
\begin{equation}
\Delta E \propto r_p^{-\alpha}.
\end{equation}
For the critical velocity at infinity, we then have
\begin{equation}
v \propto r_p^{-\alpha/2}.
\end{equation}
From Eq. (\ref{eqn3:5}), we obtain
\begin{equation}\label{eqn3:9}
v \propto \rho^{-\alpha \over \alpha + 1}.
\end{equation}
The slope of this relation between $\rho$ and $v$ lies
in the range $\{-1,0\}$, for {\it any} positive value of $\alpha$.
This implies that even large differences in the relations
between nearest approach and energy loss are not very well visible in
Fig. 5.

From a practical point of view, this reasoning has an interesting
consequence: Eq. (\ref{eqn3:9}) implies that the merging cross section is
relatively insensitive to the fact that the amount of energy loss
varies widely among different models, when measured at the same
pericenter distance.

From a theoretical point of view, we already saw that the impulsive
tidal approximation is not reasonable.  More specifically, the
time scale of the encounter is proportional to $r_p^{1.5}$. If this
encounter time scale significantly exceeds the orbital time scale of a
particle in the galaxy, the binding energy of that particle becomes an
adiabatic invariant.  As a result, the energy change decays
exponentially as a function of increasing $r_p$.  If a galaxy has a
finite radius $r$, encounters with $r_p \gg r$
will result in only a very small energy change, and $v(r_p)$ would
thus decrease exponentially.

\subsection{Theoretical Scaling Arguments}

A simple alternative to the impulsive approach could start with a
determination of the amount of significant particle overlap during the
encounter.  A quick inspection of Figs. 5 and 6 already shows that
this may be a reasonable approach, when we realize the exceptional
nature of the Hernquist and King ($W_c=1$) models.  These are the two
models that deviate most in both figures, and indeed, these are the
two models that deviate most in the position of the radius that encloses
95\% of the mass in each model, with the King ($W_c=1$) model having
an unusually sharp cut-off and the the Hernquist model having an
unusually large amount of matter far from the core region (see also
Fig. 12).

If a galaxy has an extended halo with a power-low density profile, the
particles in the region outside the radius $r_p$ have orbital
time scales comparable to, or longer than, the time scale of the
encounter.  As a result, those particles are strongly perturbed, and
we can still expect the impulsive approximation to give a reasonable
estimate, at least in order of magnitude.  Thus the energy gain of
particles outside the radius $r_p$ can be estimated from the typical
velocity kick $\Delta v$ they receive as
\begin{equation}\label{eqn3:10}
\Delta E_{out} \approx [M -M(r_p)] \left( \Delta v \right)^2 =
[M-M(r_p)]\left({2GM \over r_p v_p}\right)^2
\end{equation}
where $M(r)$ is the mass of a single galaxy enclosed inside the radius
$r$ (the total mass $M=1$ initially, in our units).  
The velocity $v_p$ at pericenter passage is related to $r_p$ in the
parabolic approximation by $\half v_p^2 \approx GM/r_p$.  This gives:
\begin{equation}\label{eqn3:11}
\Delta E_{out} \approx {G[M-M(r_p)] M\over r_p},
\end{equation}
Note that this energy change is equal to the binding energy of the
mass outside the radius $r_p$.  In other words, in our approximation
all particles are divided into two groups, an inner and an outer
group.  The inner particles are considered to be left undisturbed,
while the outer ones are significantly perturbed during the encounter.

For galaxy models with a halo described by a power-low density
$\rho_d = r^{-\beta}$, we have $[M-M(r)]\propto r^{-\beta+3}$ and
\begin{eqnarray}\label{eqn3:12}
\Delta E_{out} &\propto& r_p^{2-\beta},\nonumber\\
v &\propto& r_p^{1-\beta/2}.
\end{eqnarray}
For the Hernquist model, $\beta=4$. Therefore the power index in the
latter relation is $-1$.  For the Plummer model, $\beta=5$ and the index
becomes $-3/2$.

Fig. 7 shows the critical velocity obtained by $N$-body simulations
and that estimated using Eq. (\ref{eqn3:11}). For both Plummer and Hernquist
models, the theoretical estimate and the numerical result show a
reasonable agreement, well within a factor two for $r_p$ values in the
range $\{1,20\}$.  This is a very satisfactory result, given the
simple nature of our approximations.

\subsection{Cross Sections}

From the $\rho_m(v)$ results, plotted in Figs. 4 and 5, it is
straightforward to determine the cross section $\sigma(v)$ for mergers
to occur:
\begin{equation}
\sigma(v) = \pi \rho^2.
\end{equation}
The results are plotted in Fig. 8.

A more useful way to display these cross section is by multiplying
them with a factor $v^3$.  The first two factors of $v$ reflect the
three-dimensional nature of a Maxwellian velocity distribution, with a 
velocity space volume factor of $v^2 dv$ (see Eq. \ref{eqn4:4}), while the
third factor indicates the fact that, for a fixed target size, the
rate of merging encounters is proportional to the relative velocity
of the galaxies involved.

The results, in the form of $v^3\sigma(v)$, are plotted in Fig. 9.
There is a slight gap at the left-hand side of the curves, where it
would have been too time consuming to determine the critical merger
velocities through numerical simulations.  Fortunately, we can use the 
arguments developed in \S 3.3 to extrapolate our numerical results
leftwards of where the curves end.  For galaxy models with a halo
described by a power-low density $\rho_d = r^{-\beta}$, we find from
Eqs. (\ref{eqn3:5}) and (\ref{eqn3:12}) that
\begin{equation}
\rho \propto v^{(\beta-1)/(2-\beta)}.
\end{equation}
This leads directly to
\begin{equation}
\sigma v^3 \propto \rho^2 v^3 \propto v^{(\beta-4)/(\beta-2)}.
\end{equation}
For a Hernquist model, with $\beta=4$, we see that this quantity tends
to a constant value for vanishing $v$, while for a Plummer model this
quantity tends to scale as $v^{1/3}$ for low $v$ values.  In the rate
determinations of \S 4, we will use these analytic extrapolations to
augment the numerical data.

\subsection{ Error Discussion}

In our simulations, the total number of particles per galaxy has been
typically $N=2\times 10^3$.  Depending on the exact definition of
half-mass relaxation time $t_{hr}$ and half-mass crossing time
$t_{hc}$, we can get somewhat different relationships between these
two quantities.  Let us take as a reasonably accurate choice the
expression $t_{hr}/t_{hc} = 0.1 N/\ln(0.4N)$ (see \cite{Spitzer1987} for
the factor 0.4).  For $N=2\times 10^3$, this gives us $t_{hr}/t_{hc}
\simeq 30$.  However, we have used a softening length of 1/32.  With a
strong-deflection distance of order $1/N$, the added softening
increases the relaxation time by a factor $\log N /\log32 \simeq 2$.
Our galaxies can thus be expected to show significant relaxation after
60 half-mass crossing times.  In our units, the internal
three-dimensional velocity dispersion of a galaxy is $1/\sqrt2$, and
the time to cross the system, starting at the half-mass radius, is
therefore $\sim 2/\sqrt2 \simeq 3$.  Thus, we can expect significant
relaxation to occur after $\sim 200$ time units.

For the simulation in fig. 3, relaxation effects are not expected to
be very important, given a total duration of the simulation of 100
time units.  However, for larger impact parameters, the duration of a
typical encounter simulation can be significantly longer, and
relaxation effects may begin to influence the measurements of energy
dissipation.  This in turn will effect the determination of the border
line for mergers to take place at high values for the impact parameter.

We have run various tests in order to determine the systematic errors
that can occur for large $r_p$ values.  For example, for the King
models with $W_c=1$, we have rerun a number of experiments, using a
total particle number per galaxy in the range $512 \le N \le 16384$,
as compared to our standard value $N=2048$.  For relatively small
pericenter values, such as $r_p =0.1$ and $r_p =3.2$, we did not find
any noticeable change for the final merger boundary value $v_m(r_p)$.
In fact, even downgrading our runs by a factor four in total particle
number did not have a significant effect on the outcome.

However, for $r_p =6.4$, we found that the measured $v_m(r_p)$
steadily decreased for increasing particle numbers, until leveling off
for a value of $N$ four times larger than originally used.  As a
consequence, the reduction in slope in the lower right corner in the
line labeled {\sl King ($W_c=1$)} in Fig. 6 is not correct.  In the
limit of $N \rightarrow \infty$, the line will continue to descend
with a near-constant slope, leading to $v_m$ values around $r_p = 8$
that are roughly half of what we have obtained here for our standard
value of $N=2048$ particles per galaxy.  However, these systematic
errors do not contribute significantly to the overall reaction rates,
as we can see in Fig. 9.

\section{Merger Rates}

Given the cross section determined in the previous section, we are now 
in a position to determine the rate at which mergers occur, by
averaging over a distribution of velocities.  

The distribution function for the velocity of the galaxies inside a
cluster is given in a Maxwellian approximation by
\begin{equation}
f_1(v) = \left( {2\over\pi} \right)^{1/2}\sigma_e^{-3}
v^2 e^{-v^2/2\sigma_e^2},
\end{equation}
where $\sigma_e$ is the one-dimensional velocity dispersion for the
cluster.

The distribution of encounter velocities is then given by
\begin{equation}
f_2(v) = 2^{-1}\pi^{-1/2}\sigma_e^{-3}
v^2 e^{-v^2/4\sigma_e^2},
\end{equation}
where in both cases we use a normalization condition of unity after
integration over velocities:
\begin{equation}
\int_0^\infty f_1(v)dv = \int_0^\infty f_2(v)dv = 1.
\end{equation}

Let us introduce the ratio $x$ of the one-dimensional velocity dispersion of
the cluster and the one-dimensional internal velocity distribution of
the stars in each galaxy:
\begin{equation}
x = {\sigma_e \over \sigma_i} = \sqrt{6}\sigma_e,
\label{eq:x}
\end{equation}
since ${\sigma_i} = 1/\sqrt{6}$ in our units in which the specific
kinetic energy ${3\over 2} \sigma_i^2 = 1/4$.
We thus have
\begin{equation}\label{eqn4:4}
f_2(v) = 2^{-1}\pi^{-1/2}\sigma_e^{-3}
v^2 e^{-{3\over2}v^2/x^2},
\end{equation}

The number of merging events per unit time per unit volume is given by
\begin{equation}\label{eqn4:5}
R(n,x) = n^2\int_{v=0}^{\infty} v f_2(v)\sigma(v) dv,
\end{equation}
where $n$ is the number density of the galaxies and $\sigma(v)$ is the
cross section of merging.

We can also write this result as
\begin{equation}\label{eqn4:6}
R = n^2 2^{-1}\pi^{-1/2}\sigma_e^{-3} R(x),
\end{equation}
where $R(x)$ is the non-dimensional merging rate defined as
\begin{equation}\label{eqn4:7}
R(x) = \int_0^{v_{crit}} v^3\sigma(v)e^{-{3\over2}v^2/x^2} dv,
\end{equation}
where the upper limit of integration is reached at $v_{crit}$, since
the cross section $\sigma(v) = 0$ for $v > v_{crit}$.

Note that in our units it is not immediately clear that this last
expression is dimensionless.  If we remind ourselves of the fact that
our internal one-dimensional velocity dispersion ${\sigma_i} =
1/\sqrt{6}$ and the virial radius of an individual galaxy $r_v = 1$,
we can rewrite the expression for $R(n,x)$ as
\begin{equation}\label{eqn4:8}
R = n^2 2^{-1}\pi^{-1/2}r_v^2 {(36\sigma_i^4)\over\sigma_e^3}R(x)
= {18 \over \sqrt\pi} {1\over x^3} n^2 r_v^2 \sigma_i R(x),
\end{equation}
from which it is clear that $R(x)$ is dimensionless, since all
physical factors (density squared $\times$ cross section $\times$ velocity)
have been displayed here explicitly.

In rich clusters, the velocity dispersion $\sigma_{e}$ in the cluster
is several times larger than the internal velocity dispersion
$\sigma_{i}$ in each of the galaxies.  Even for less rich clusters,
$\sigma_{e} > \sigma_{i}$, typically.

Thus, for most rich clusters, the merging rate calculated with
constant $f(v)$ gives an error less than a few percent.  For that
regime, the dimensionless merging rate is given by
\begin{equation}\label{eqn4:9}
R_\infty = \lim_{x \rightarrow \infty} R(x) =
\int_0^{v_{crit}} v^3\sigma(v)dv.
\end{equation}

Note that $R_\infty$ does not depend on the velocity distribution of
galaxies within the cluster.  The only dependency on the external
velocity dispersion $\sigma_i$ is through the factor $\sigma_e^{-3}$
in the expression for $R$ itself.  This is simply the dilution factor
in velocity space: increasing the external velocity dispersion
decreases the density of galaxies within the region in which
encounters can lead to merging.

Table 1 shows the nondimensional merging rate $R_\infty$ for all models. 
For non-zero values of $x$, we can approximate $R(x)$ for most models
by the following fitting formula
\begin{equation}\label{eqn4:10}
R_P(x)= {12 x^2 \over x^2 + 0.42},
\end{equation}
while for the Hernquist model a better approximation is given by the
fitting formula
\begin{equation}\label{eqn4:11}
R_H(x)= {13.8 x^2 \over x^2 + 0.28},
\end{equation}
The physical reason behind the fact that the Hernquist model shows a
merging rate that is 15\% larger than the Plummer model lies in the
more extended mass distribution of the former, as is illustrated in
Fig. 12.

\section{Tidal effects}
In our discussion of the merger cross section and the merger rate, we
have so far regarded two galaxies as actually merging if the orbital
binding energy of the two galaxies after the first encounter is
negative. Simple and straightforward as this assumption is, it may not
be realistic in some extreme cases, as for example in rich clusters.
In such cases, even if two galaxies have become formally bound after
their first encounter, they may still be disrupted under the influence
of either the tidal forces from the cluster as a whole, or the tidal
forces from individual nearby galaxies. In the following, we evaluate
the magnitude and relevance of both of these effects.

\subsection{Tidal effects from the cluster as a whole}

Consider a cluster consisting of $N$ galaxies.  The tidal force from
the cluster would dissolve a pair of galaxy if the separation of the
pair is larger than $fR(ND)^{-1/3}$, where $R$ is the radius of the
cluster and $D$ is the ratio of the total mass of the cluster to the
mass associated with the individual galaxies.  The coefficient in this
expression would be $f=1/3$ in the limit in which the cluster mass is
all concentrated in the center (Spitzer, 1987, eq. 5-5).  For more
realistic mass distributions, the tidal force on a typical merging
pair is considerably less, and therefore $f$ is significantly larger.
A reasonable approximation is therefore to simply take $f=1$.

Thus, we discard a candidate merger pair if their first apocenter distance
exceeds
\begin{equation}
r_{t,global}  = R(ND)^{-1/3}.
\label{eq:aparent}
\end{equation}

\subsection{Tidal effects from individual nearby galaxies}

To determine the tidal effects of neighboring galaxies, acting as
perturbers, requires a somewhat more complicated calculation.  For a
pair of galaxies in the process of merging to become unbound by the
encounter with a third galaxy, this latter galaxy should pass within a
distance that is sufficiently close, during a time that is
sufficiently quick, {\it i.e.} before the first fall-back.  Since most
orbits after the first fly-by are highly eccentric, the fall-back time
can be approximated by the orbital period itself, which is given
for a pair of galaxies by
\begin{equation}
T = \pi\sqrt{2a^3 \over m},
\end{equation}
where $a$ is the semi-major axis of the pair and $m$ is the mass of
one galaxy.

We can apply the usual $n\sigma v$ argument for the rate, to determine a
disruption criterion of the pair. For the pair to be disrupted, we have
\begin{equation}
n\sigma v T \ge 1,
\label{eq:nsigmav}
\end{equation}
as an approximate formula. Here we can estimate the density $n$ to be
of order
\begin{equation}
n = N / R^3.
\end{equation}
A typical relative velocity value is given by the virial theorem as
\begin{equation}
v \sim \sqrt{2DNm\over R},
\label{eq:ourv}
\end{equation}

We neglect gravitational focusing for the perturbers, which
will fly by with a velocity much higher than the relative velocity of
the two merger candidates around the apocenter of their first orbit.
This means that we can use the geometrical cross section
\begin{equation}
\sigma = \pi r^2
\end{equation}
We can make a quantitative estimate for this cross section as 
follows. The average amount of energy transferred by a third galaxy
which passes at a
distance $r$ from one of the galaxies in the pair is given in the
impulse approximation by
\begin{equation}
\Delta E \sim {1 \over 2} m \left({m \over v r}\right)^2.
\end{equation}
This energy change must be larger than the binding energy of the pair
$m^2/2a$. Thus, we obtain
\begin{equation}
r^2 \sim am/v^2
\end{equation}
as the maximum encounter distance to disrupt a pair. 

We can find a second expression for this distance, using equation
\ref{eq:nsigmav}.  Using both expressions to eliminate $r$,
substituting $v$ from equation (\ref{eq:ourv}), we find
\begin{equation}
a > \left(2/\pi^4\right)^{1/5}RN^{-1/5}D^{1/5},
\end{equation}
as the condition for disruption.  Taking into account that the first
bound orbit, after the initial encounter, is likely to be highly
eccentric, we can approximate the apocenter distance after the first
encounter to be $r_{apo} \simeq 2a$.  The criterion for tidal disruption from
a single nearby galaxy then becomes $r_{apo} > r_{t,local}$ with
\begin{equation}
r_{t,local} = RN^{-1/5}D^{1/5}.
\label{eq:athird}
\end{equation}

Comparing eqs. \ref{eq:aparent} and \ref{eq:athird}, we see that the
latter critical distance exceeds the former by a factor 
\begin{equation}
r_{t,local} / r_{t,global} =  N^{2/15} D^{8/15} > 1,
\label{eq:athird2}
\end{equation}
since $D \ge 1$ by definition.  Typical values for this ratio will be
$\sim10$ for rich clusters.  

In conclusion, we see that the tidal forces exerted by neighboring
galaxies are less important than the tidal force exerted by the parent
cluster as a whole, in its effectiveness to disrupt a galaxy pair
while it has become bound after its first encounter.

How large is the critical distance $r_t = \min(r_{t,local}, r_{t,global})
= r_{t,global}$ for tidal disruption in real clusters? For a cluster of
$N$ galaxies, and using the notion introduced in eq. (\ref{eq:x}), we
find that the half-mass radius of the cluster is larger than that of
the individual galaxies by a factor of $ND/x^2$. Thus, using
eq. (\ref{eq:aparent}) we obtain
\begin{equation}
r_t = (ND)^{2/3}x^{-2}r_h,
\end{equation}
for the maximum apocenter distance after the encounter at which tidal
disruption can be avoided.

To sum up, tidal effects will disrupt those mergers that have a
binding energy less than a critical value $E_t$, given by
\begin{equation}
E_t \sim {M^2 \over r_t} =
\displaystyle{-5x^2E \over (ND)^{2/3}} =
\displaystyle{{5\over4}{x^2 \over (ND)^{2/3}}}
\label{eqnec}
\end{equation}
where we have used equation \ref{somethingwith0.8}, with $E$ the
internal energy of one galaxy ($E=-1/4$ in our units, see
equation (\ref{somethingwith0.25})).

For typical clusters, $E_t$ lies in the range $0.01 \simlt E_t \simlt 0.1$.
For example, for a cluster with $N=100$, $D=3$, and $x=2$, we have
$E_t \sim 0.1 $, while for a rich cluster with $N=1000$, $D=10$, and
$x=3$, we have $E_t \sim 0.02$.

\subsection{Sensitivity of cross sections to tidal effects}

In order to calculate the cross section as a function of critical
binding energy for tidal disruption $E_t$, we could perform a
systematic search as we did for the case of $E_t = 0$. Since this
would require a rerun of many of our simulations, separately for each
value of $E_t$, it would be preferable to check whether an appropriate
approximation exists that may allow us to derive a merging criterion
for non-zero $E_t$ to a reasonable accuracy.  

The merging criterion is expressed as
\begin{equation}
E_1 = E_0 + \Delta E < E_t,
\end{equation}
where $E_0$ and $E_1$ are the orbital binding energy of two galaxies
before and after the first encounter, and $\Delta E$ is defined as
$E_1 - E_0$.  $\Delta E$ is a function of collision parameters $\rho$
and $v$, or $r_p$ and $v_p$, as described in section 3.

As an approximation, we try the assumption that $\Delta E$ is
determined essentially by $r_p$ alone, at least when the relative
velocity is close to the critical velocity for merging. Thus, we have
in first approximation:
\begin{equation}
\Delta E = - \displaystyle{1 \over 2} \mu v_{crit}^2,
\end{equation}
and therefore 
\begin{equation}
E_1 =  \displaystyle{1 \over 2}\mu (v^2 -v_{crit}^2).
\label{detheory}
\end{equation}
It is reasonable to expact this assumption to be quite good for most
cases since $v_p$ is determined mainly by the potential energy at
minimum separation, and depends only weakly on $v$, the velocity at infinity.

To test whether this approximation is satisfactory, we compare the result of 
our numerical experiments and the theoretical estimate obtained
assuming  that $\Delta E$ is independent of the velocity at
infinity. Figure 13 shows the orbital energy of the relative motion of
the two galaxies, in units of the kinetic energy associated with the
critical velocity $v_{crit}$
\begin{equation}
\epsilon_{orb} = \displaystyle{ 2 E_{orb} \over \mu v_{crit}^2},
\label{eqforfig12bwhichisreallyfig13}
\end{equation}
plotted against the initial orbital velocity at infinity, in units of
the critical velocity. Note that $v_{crit}$ is calculated
for the pericenter distance $r_{p}$ derived from the impact parameter
$\rho$ through the two-body Keplerian gravitational focusing
approximation. The approximation of velocity-independent energy
dissipation, given by equation (\ref{detheory}), is given here as the
solid curve.

The agreement between theory and experiment is excellent for a
wide range of initial relative velocities. Therefore, in the
following we use the assumption that $\Delta E$ is a function of
$r_p$ only. 

Figure 14 shows the effects of non-zero $E_t$ in terms on the
differential merger rate $v^3\sigma$.  It is clear that the reduction
in the merging probability is significant, especially for the
Hernquist model.  The reduction is largest for lower velocities, for
which a typical binding energy of the orbit after the first encounter
is lower, and therefore the vulnerability to tidal effects larger.

Figure 15 gives the nondimensional merger rate $R = R(\infty)$ as a
function of $E_t$. The dependence of $R_{\infty}$ on $E_t$ is similar
for different models. For rich clusters with $E_t \sim 0.02$, the
merger rate is $3/4 \sim 1/2$ of the value for $E_t = 0$. For a more
modest cluster, with $E_t \sim 0.08$, the merger rate has dropped to
$1/3 \sim 1/5$ of the value for $E_t = 0$.

Theoretically, $R$ should reach zero for some 
finite value of $E_t$.  Thus, a fitting formula of the form
\begin{equation}
R = R_0 \displaystyle{\left({E_{t,0} - E_t \over E_{t,0}}\right)}^{\gamma}
\end{equation}
would be an appropriate formula.  We found $E_{t,0} = 0.25$ and $\gamma=3$
to give a reasonable fit for both a King model ($W_c=7$) and a Plummer model.

\section{Applications: Merging in Clusters of Galaxies}

As a straightforward application of the merger rates we have
determined in the preceding sections, let us take a cluster of
galaxies with a 3-D spatial density $n$, and a 1-D velocity dispersion 
$\sigma_e$ for the motion of the galaxies in the cluster.  These
galaxies are all considered to be given by identical Hernquist models, 
with half-mass radius $r_h$ and internal 1-D velocity dispersion $\sigma_i$.
We can use the asymptotic expression $R_\infty$ for the merger rate,
given by Eqs. (\ref{eqn4:9}) and (\ref{eqn4:11}), with the relation between the virial
radius $r_v$ and the half mass radius $r_h$ given by Eq. (\ref{eqn2:4}).

For a cluster of galaxies, with a one-dimensional velocity dispersion
of $\sigma_{e}$ km/s, we can express our main result in physical
units as follows.  
Under the assumption that all galaxies have
identical halos, with internal one-dimensional velocity dispersion of
$\sigma_{i}$ km/s, we find a relative merger rate (per unit volume
of $1\Mpc^3$ per Gyr) of
\begin{equation}
R_{rel} = 0.0084 \left({n\over1/\Mpc^3}\right)^2
\left({r_h\over 0.1 ~ \Mpc}\right)^2
\left({\sigma_i\over 100~ \kmps}\right)^4
\left({300~ \kmps\over \sigma_e}\right)^3
\Mpc^{-3}\Gyr^{-1}
\end{equation}
where $n$ is the density of galaxies in the galaxy cluster.  If we
approximate the cluster by a collection of $N$ galaxies evenly
distributed within a radius $R$, we find a total merger rate of 
\begin{equation}\label{eqn5:2}
R_{tot} = 0.0020 ~ N^2 \left({1~\Mpc\over R}\right)^3
\left({r_h\over0.1~\Mpc}\right)^2
\left({\sigma_i\over 100~ \kmps}\right)^4
\left({300~ \kmps\over \sigma_e}\right)^3
\Gyr^{-1}
\end{equation}
corresponding to a fiducial merger time scale of
$ t_m = 5.0 \times 10^{11} \yr $.  In order for merging to occur at a
rate that is significant over a Hubble time, we would require a rate
that is two orders of magnitudes larger.  For example, we could
take a cluster with 100 galaxies, within a radius of 1 Mpc.  If we
keep the same fiducial values for the other parameters, given in
Eq. (\ref{eqn5:2}), mergers would occur at a rate of, on average, once every 50
Myr.

This merger rate is an overestimate, however, since we have not yet
taken into account the disruptive effects of tidal interactions,
discussed in the previous section.  The corresponding correction
factor, stemming from a finite $E_t$,  is fairly large. For 
example, for $N=100$, $D=10$ and $\sigma_i/\sigma_g = 0.33$, the
correction factor $R/R_0$ is around 0.25. Thus, we conclude that in
the above case a merger occurs only once every 200 Myr.

Here we assumed that the cluster is in dynamical equilibrium, which is
certainly not true for clusters with, for example, substructures. To
apply our result to such clusters, we need rather detailed knowledge
about how the substructures evolve.

\section{Summary }

We have presented cross sections (\S3) and reaction rates (\S4) for
merging events to happen during encounters of equal-mass spherical
galaxies, first in isolation, and then in the presence of a background
cluster (\S5).  Our results are easily applied to estimate the merger rate
in clusters of galaxies (\S6).

At first glance, the restriction to spherical galaxies might seem
somewhat restrictive.  However, our numerical calculations as well as
our very successful analytic approximations show that the merger rate
is largely determined by the outer few-to-ten percent of the mass
distribution in the galaxies.  For realistic galaxies, even with only
a relatively small amount of dark matter, it is therefore the halo
mass distribution that determines the merger rate.  Even if halos were
significantly flattened, it would be rather surprising if our results
would be affected by more than a factor of two.  For practical
purposes, therefore, we are confident that our results can be applied
to galaxy clusters, directly as they are given in \S6.\


We thank the referee, Josh Barnes, for helpful comments on the
manuscript.  This work was supported in part by the Grant-in-aid for
Specially Promoted Research (04102002) of the Ministry of Education,
Science, and Culture, by the National Science Foundation under Grant
No. PHY89-04035, by National Science Foundation under an Advanced
Scientific Computing grant ASC-9612029, and by a grant from the
Ministry of Education of Japan for a summer visit of P.H. to Kyoto.
P.H. acknowledges the hospitality of Dr. Masataka Fukugita and the
Yukawa Institute for Theoretical Physics in Kyoto, where this paper
was started.  We both acknowledge the hospitality of the Institute for
Theoretical Physics in Santa Barbara, where we continued to write part
of this paper.

\clearpage

\begin{deluxetable}{cl}
\footnotesize
\tablecaption{Nondimensional Merger Rate}
\tablewidth{0pt}
\tablehead{
\colhead{Model} & \colhead{$R_\infty$}
} 
\startdata
Plummer & 11.9\nl
Plummer ($r_{cut}=4$) & 11.5\nl
King ($W_c=1$) & 12.4\nl
King ($W_c=7$) & 11.8\nl
Hernquist & 13.8\nl

\enddata
 
\end{deluxetable}

\clearpage

\figcaption[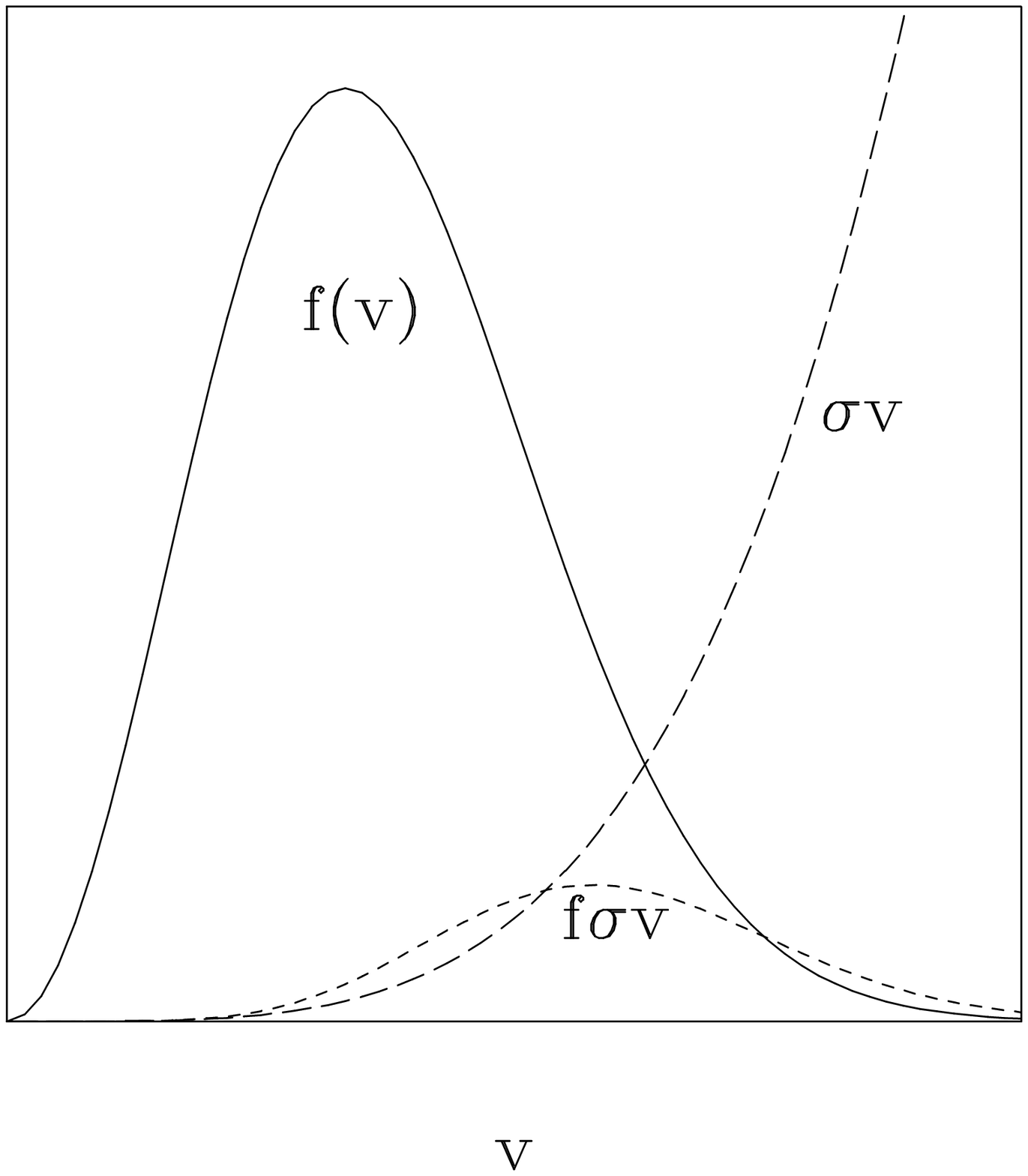,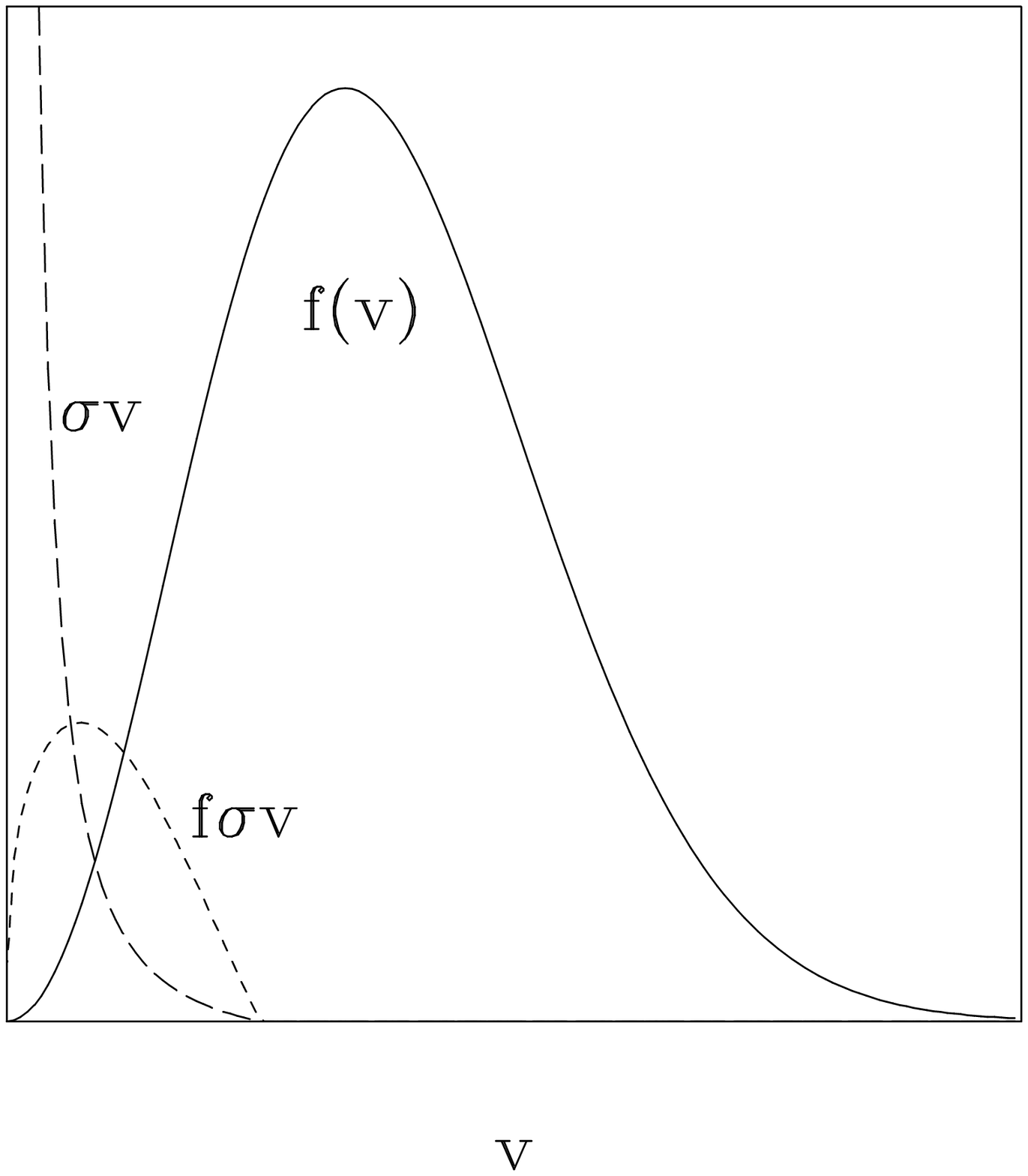]{
a) Nuclear reaction rates in the core of a star, as the product of
cross section $\sigma$ and the rate factor $vf(v)$, with $v$ the
relative velocities of the nuclei, and $f(v)$ their Maxwellian
distribution.  b) Merger rates in a rich cluster of galaxies, similar
to the stellar evolution case.
\label{fig1}}
\figcaption[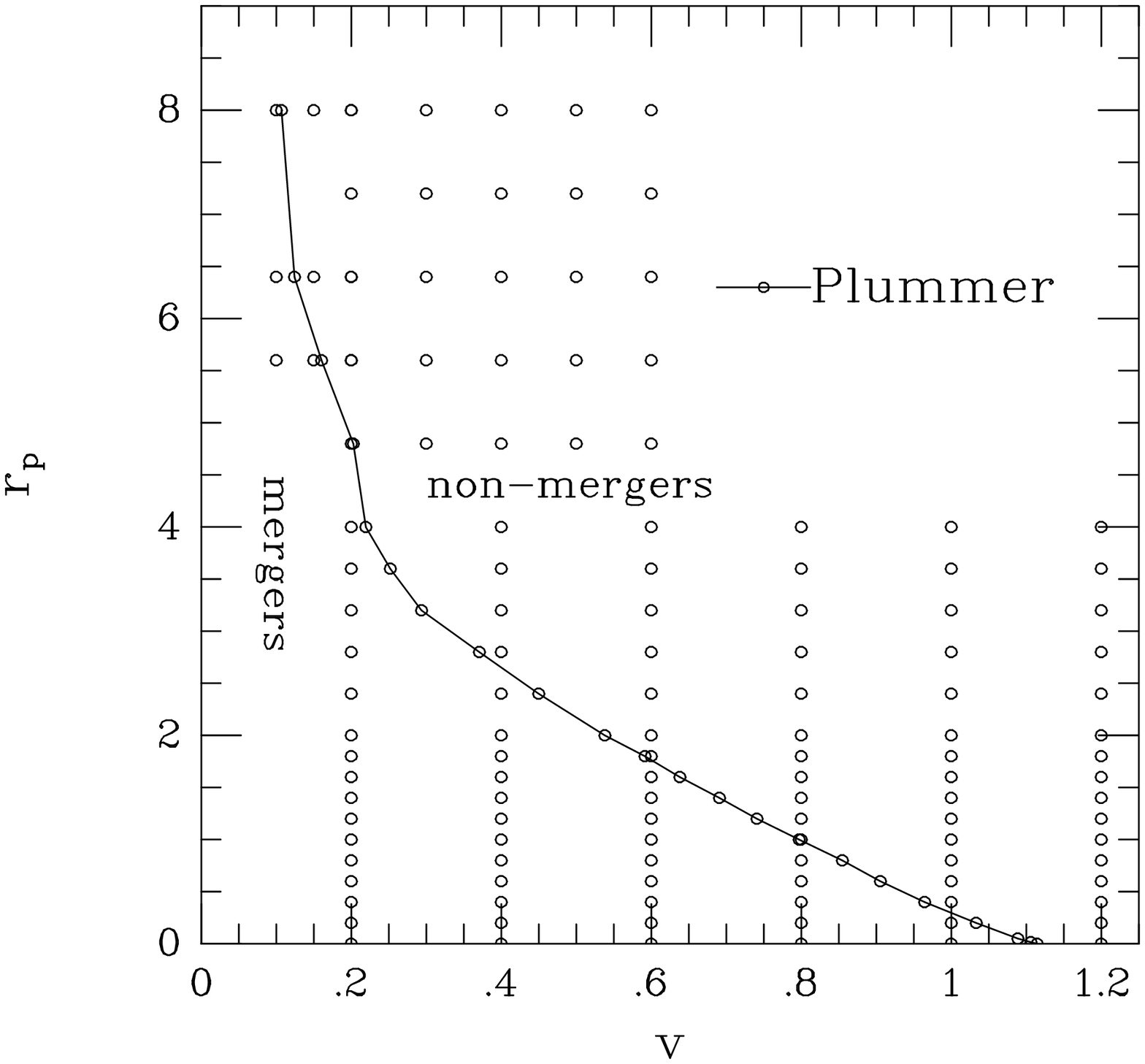]{
For each value of the pericenter distance $r_p$, circles indicate the
choice of asymptotic velocities $v$ for which a galaxy encounter
has been carried out.  Through linear interpolation of the energy
dissipation, the critical velocity $v(r_p)$ for producing
parabolic outgoing orbits is found for each $r_p$ value.  The full
line connecting these points indicates the border of the region within
which merger takes place.
\label{fig2}}

\figcaption[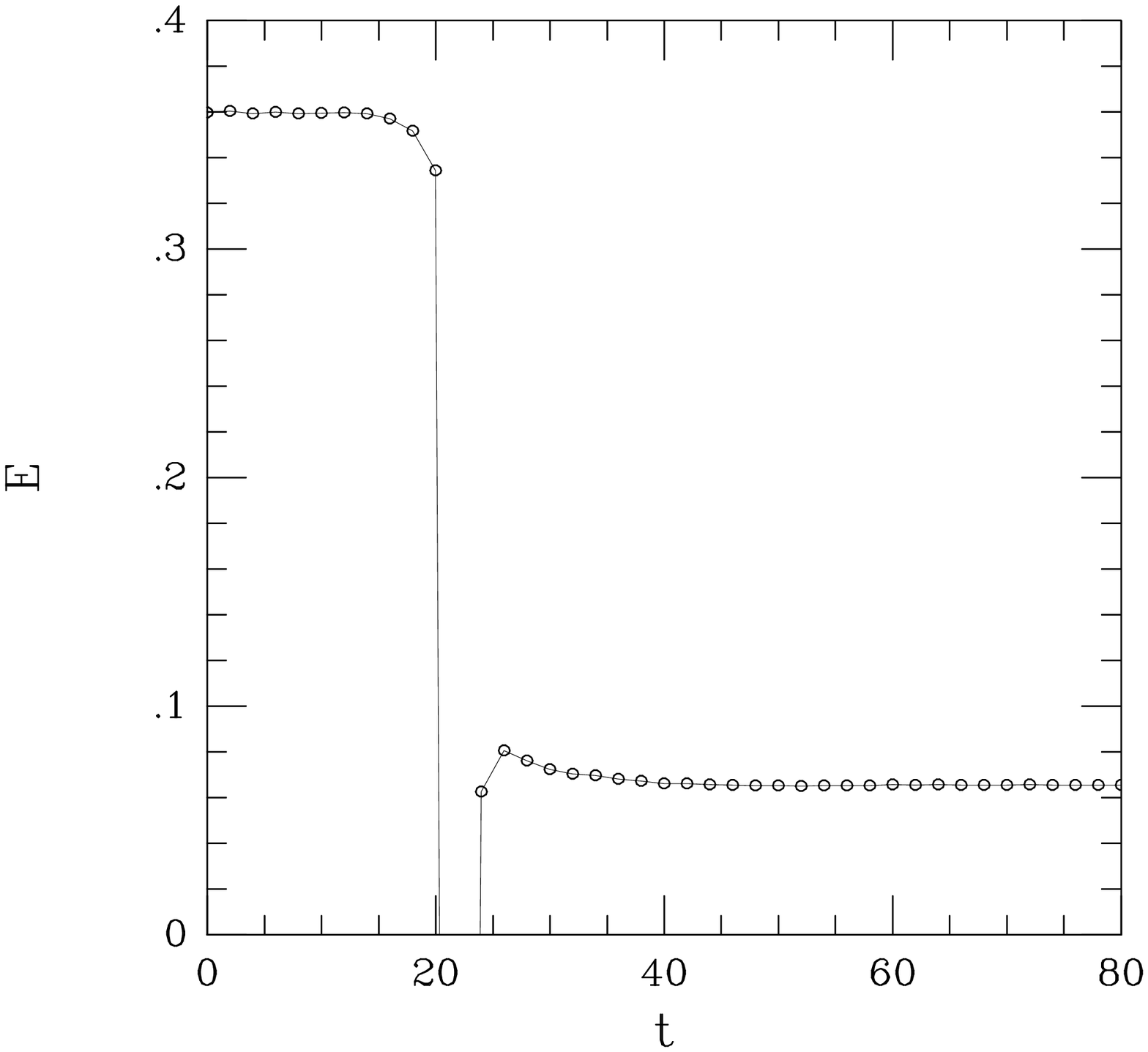]{
Total energy $E(t)$ of the two galaxies after the first encounter, as a
function of time $t$, as specified in Eq. (7).  The initial
lowering of the total energy is caused by the fact that the tidal
interaction terms have been neglected in the calculation of the
potential.
\label{fig3}}

\figcaption[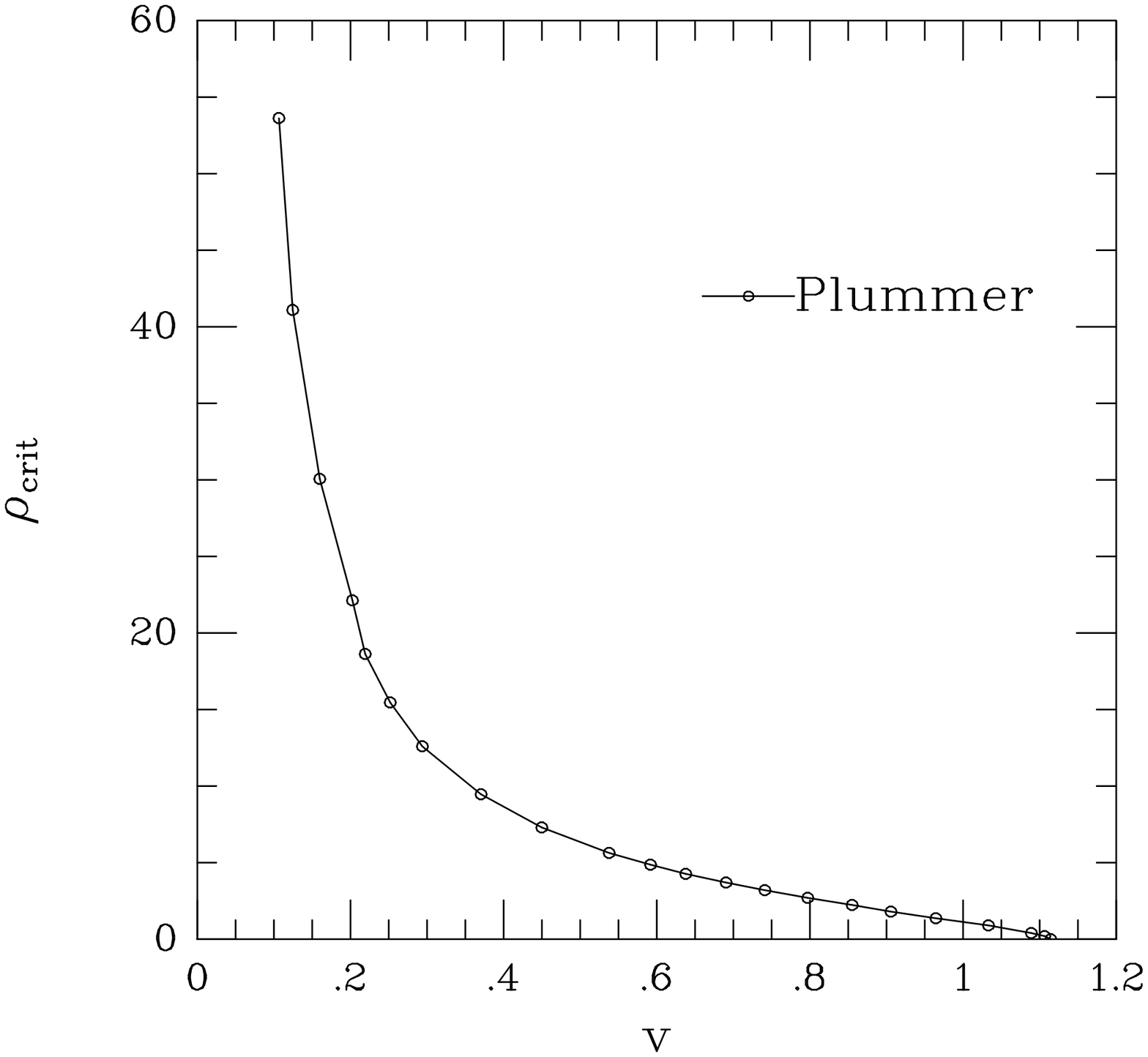,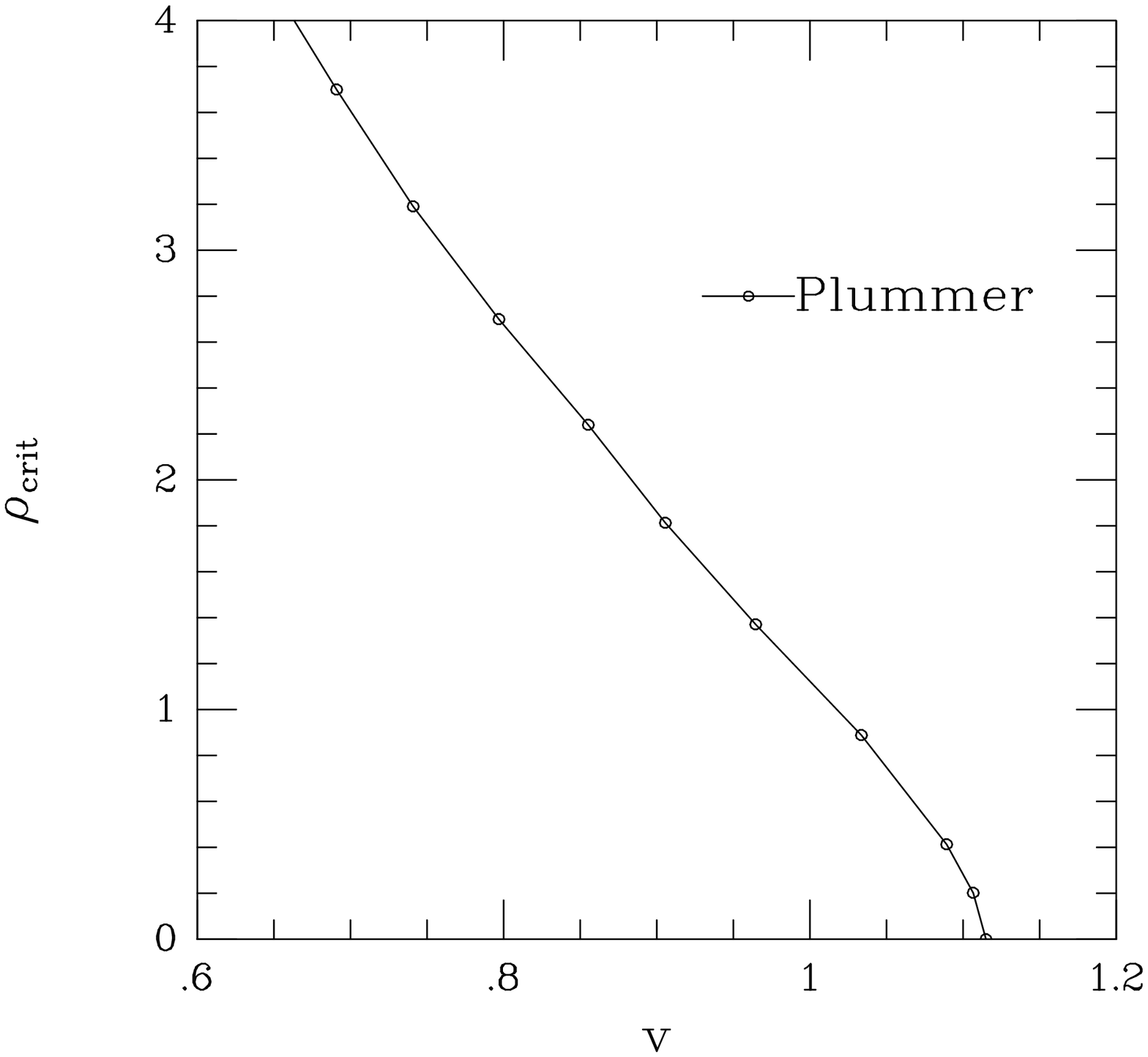]{
Maximum impact parameter $\rho_{crit}(v)$ for which merging takes
place, as a function of velocity at infinity $v$, for Plummer
model initial conditions.  Fig. 4b is an enlargement of fig. 4a, for
small impact parameter values.
\label{fig4}}

\figcaption[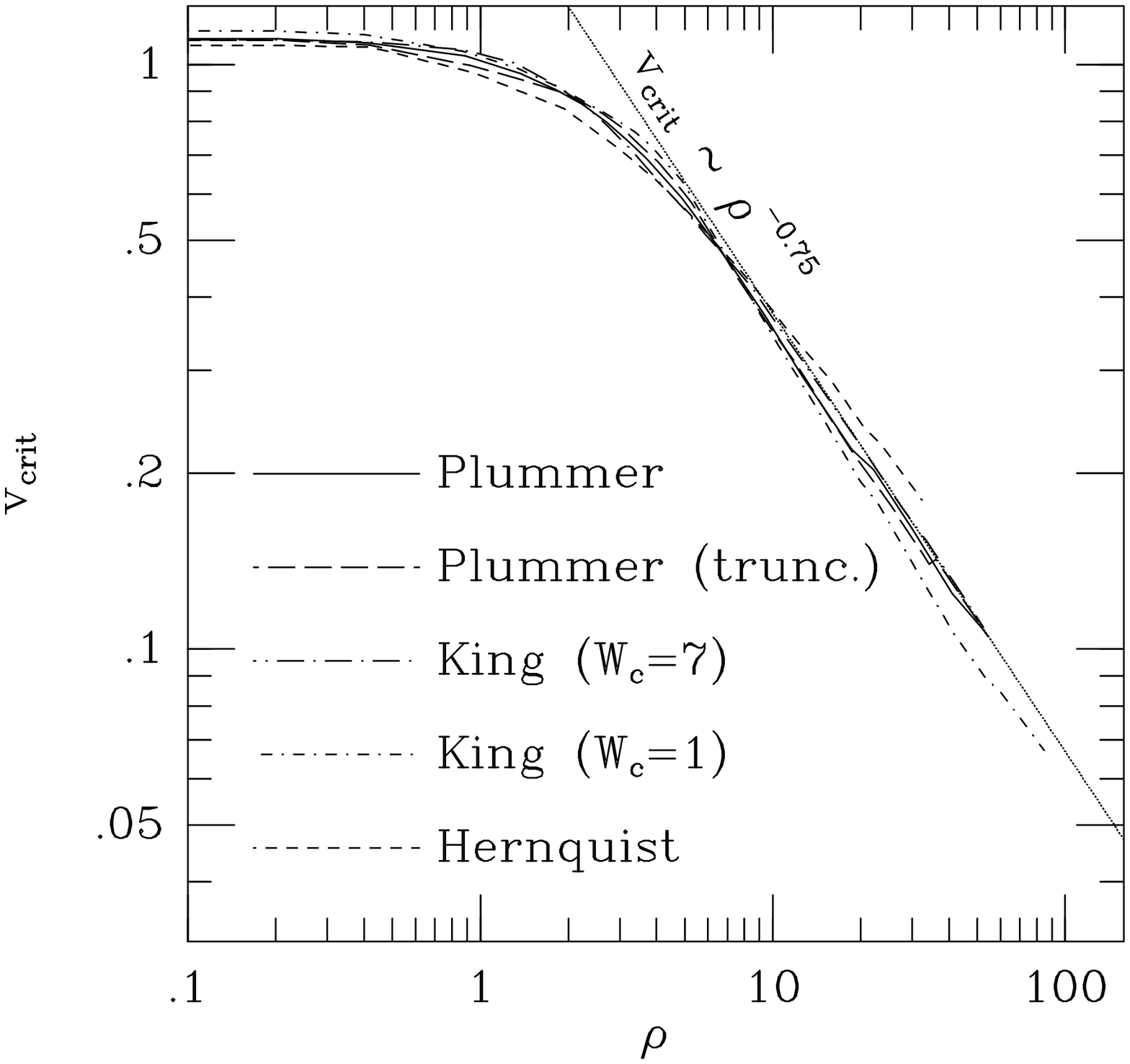]{
Maximum velocity $v_{crit}(\rho)$ for which merger takes place.  The impact
parameter $\rho$ and $v_{crit}$ are both asymptotic values, in the limit of
infinite initial galaxy separation. Solid, short dashed, long dashed,
short dot dashed and long dot dashed curves are results for Plummer
($r_{cut}=23$),
King ($W_c=1$), King ($W_c=7$), Plummer ($r_{cut}=4$), and Hernquist
models, respectively. 
\label{fig5}}

\figcaption[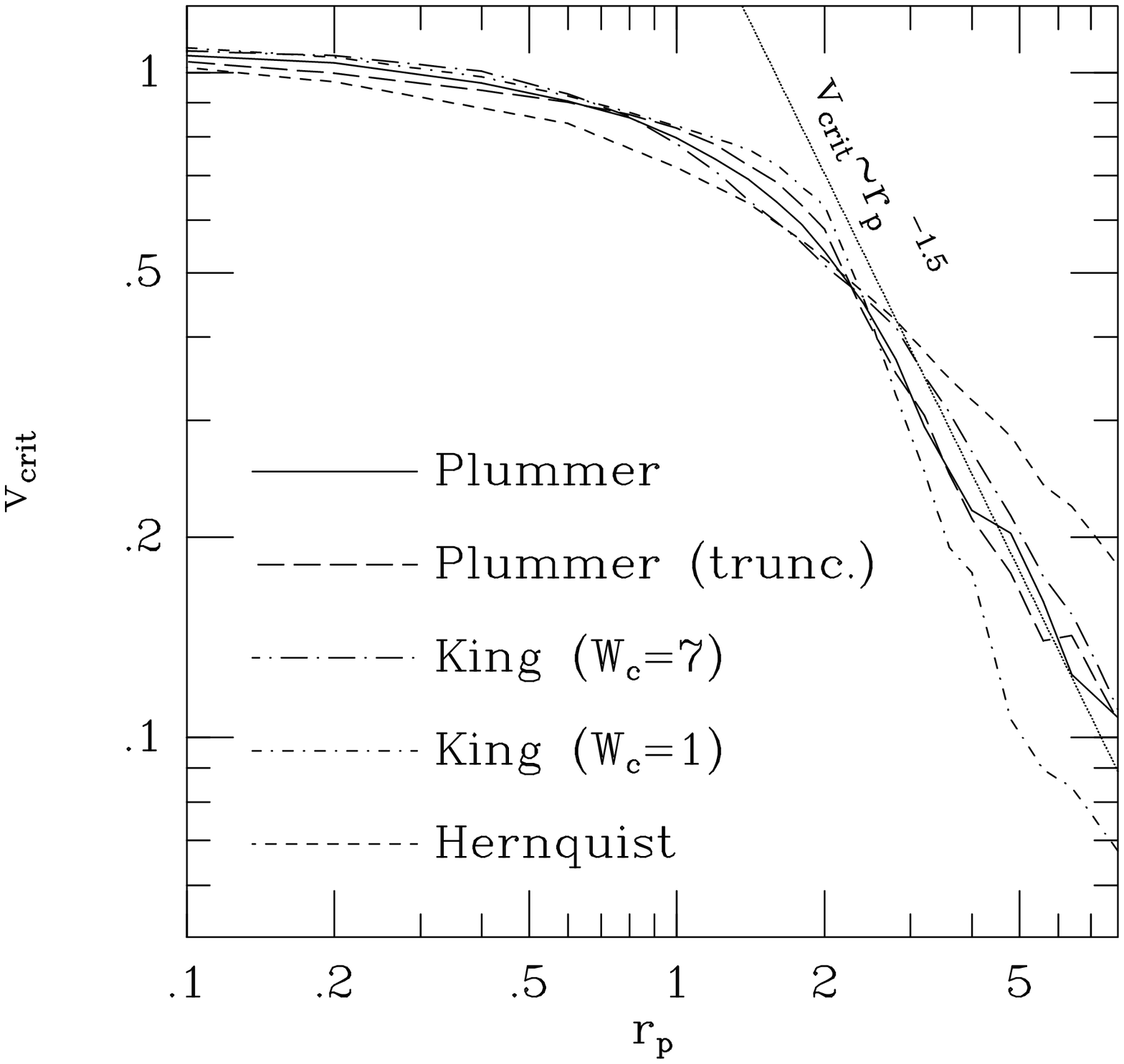]{
As figure 5 but with the merger velocity $v_{crit}(r_p)$ now plotted
against the pericenter distance $r_p$.  The tidally impulsive limit,
the line $v_{crit} \propto r_p^{-3/2}$, is given here for comparison.
\label{fig6}}

\figcaption[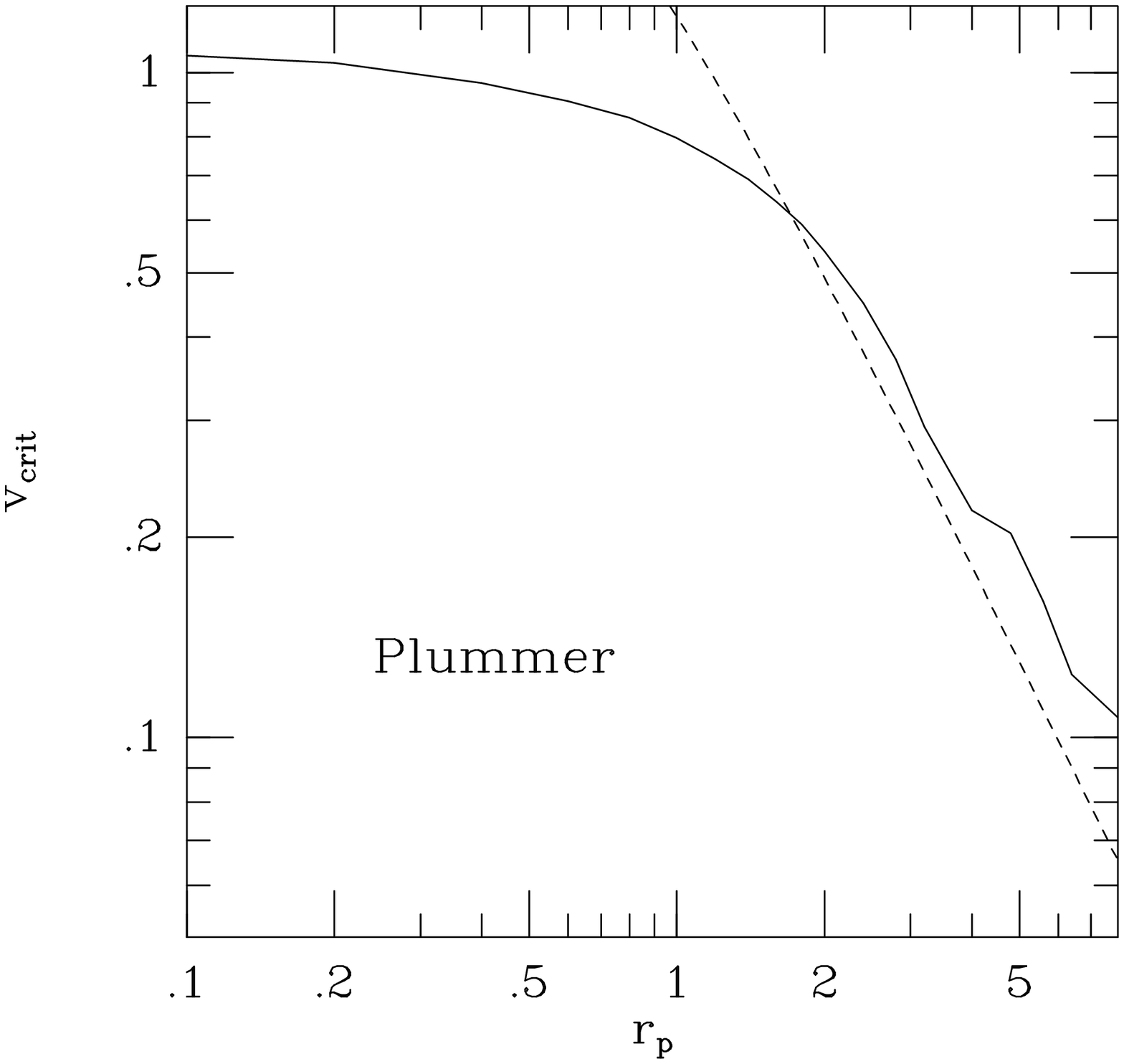,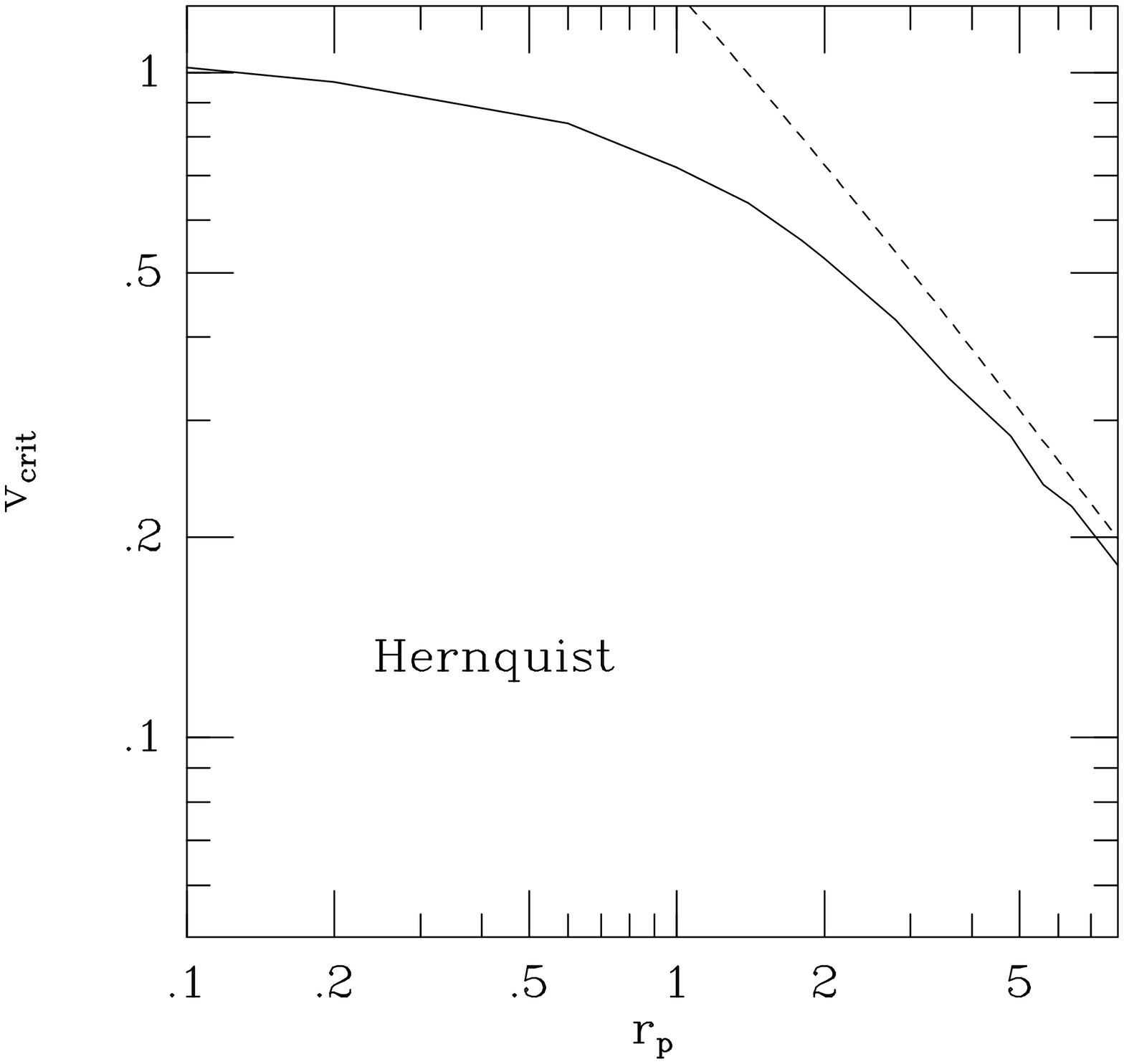]{
Same as figure 6 but including the theoretical estimate for large
$r_p$.  Fig. 7a shows the experimental results for a Plummer model,
together with the theoretical estimate $v_{crit} \propto r_p^{-3/2}$.
Fig. 7b shows similar results for a Hernquist model, together with the
theoretical estimate $v_{crit} \propto r_p^{-1}$.
\label{fig7}}

\figcaption[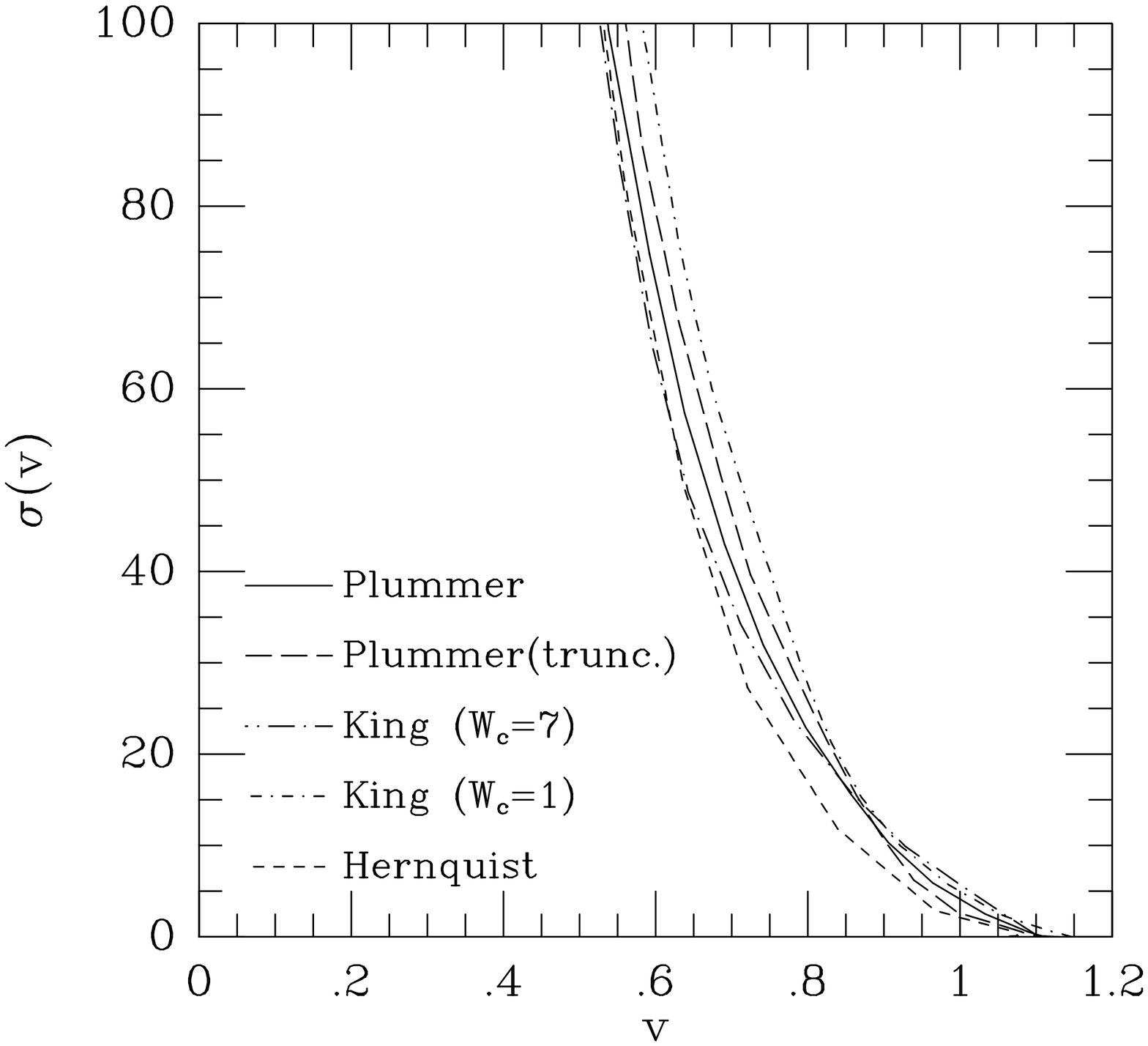]{
Merging cross section $\sigma(v)$ where $v$ is the initial encounter
velocity (at infinity).  The various lines have the same meaning as
those in Fig. 5.
\label{fig8}}

\figcaption[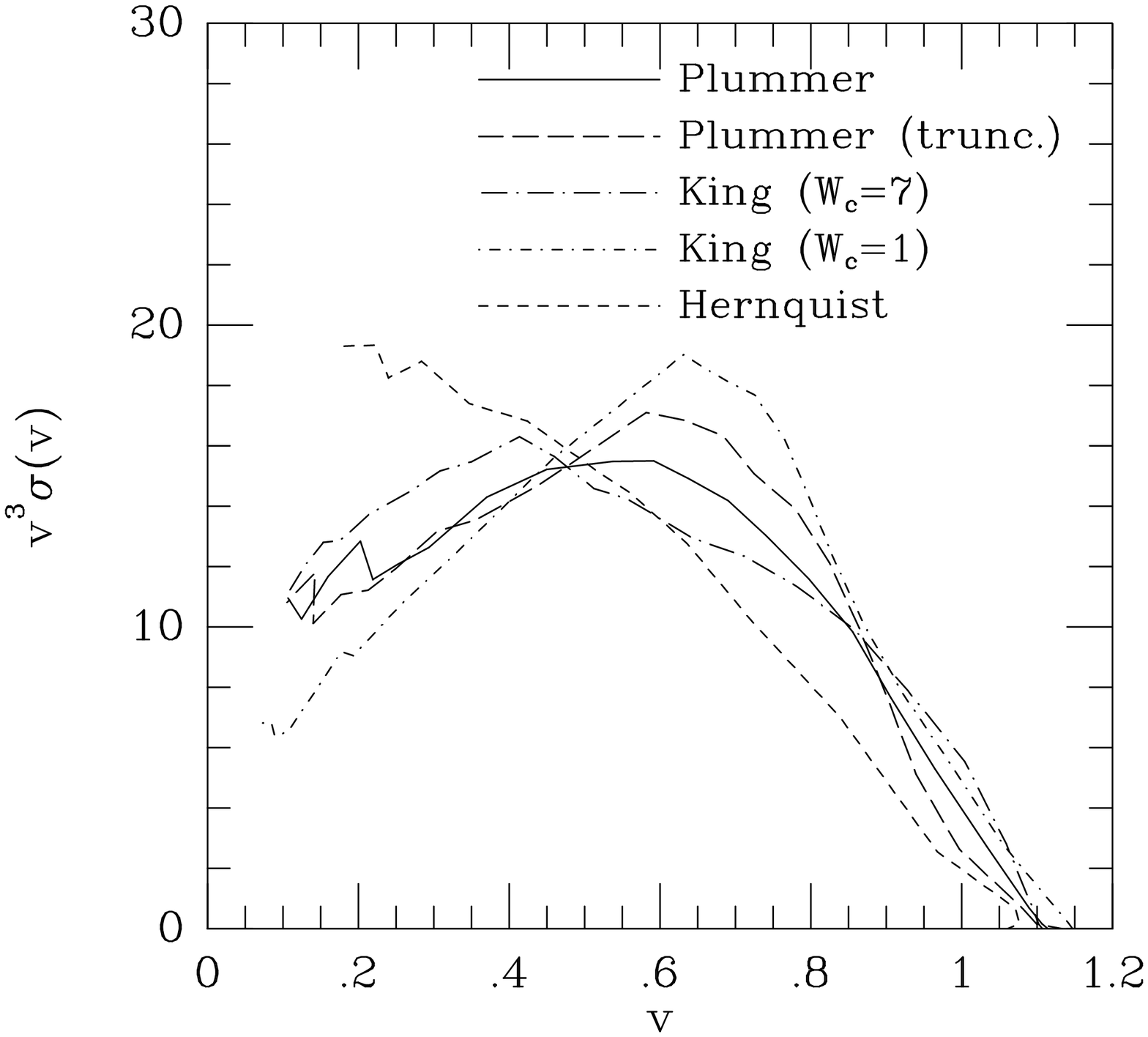]{
The differential merging rate $v^3\sigma(v)$ plotted as a function
of the velocity $v$ for five different galaxy models. The curves have
the same meanings as in figure 5.
\label{fig9}}

\figcaption[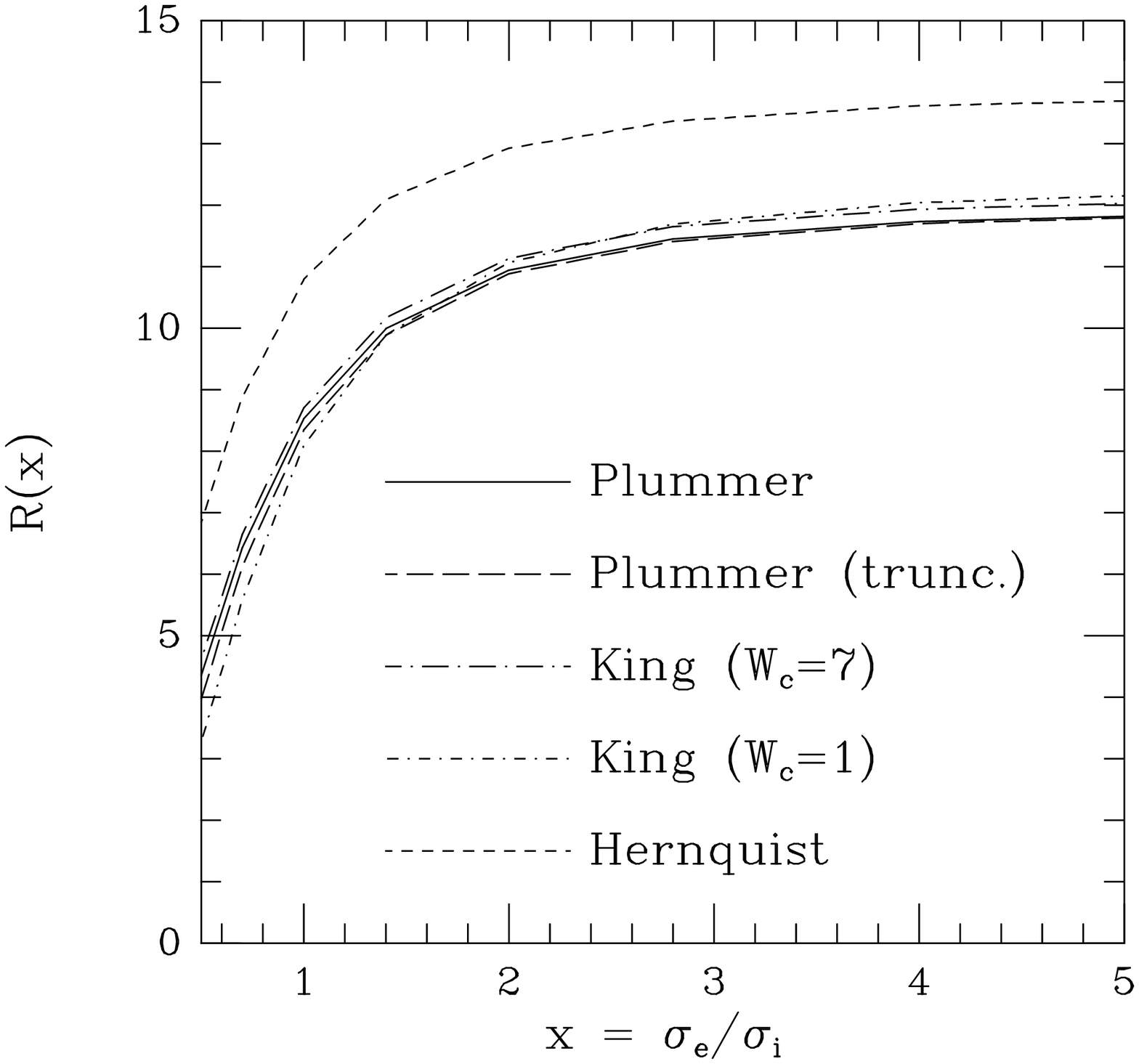]{
Merger rate $R(x)$ as a function of the ratio
of velocity dispersions: $\sigma_{e}$ indicates the velocity
dispersion of the galaxies inside the galaxy cluster, and $\sigma_{i}$
the internal velocity dispersion for the stars inside each galaxy model.
\label{fig10}}

\figcaption[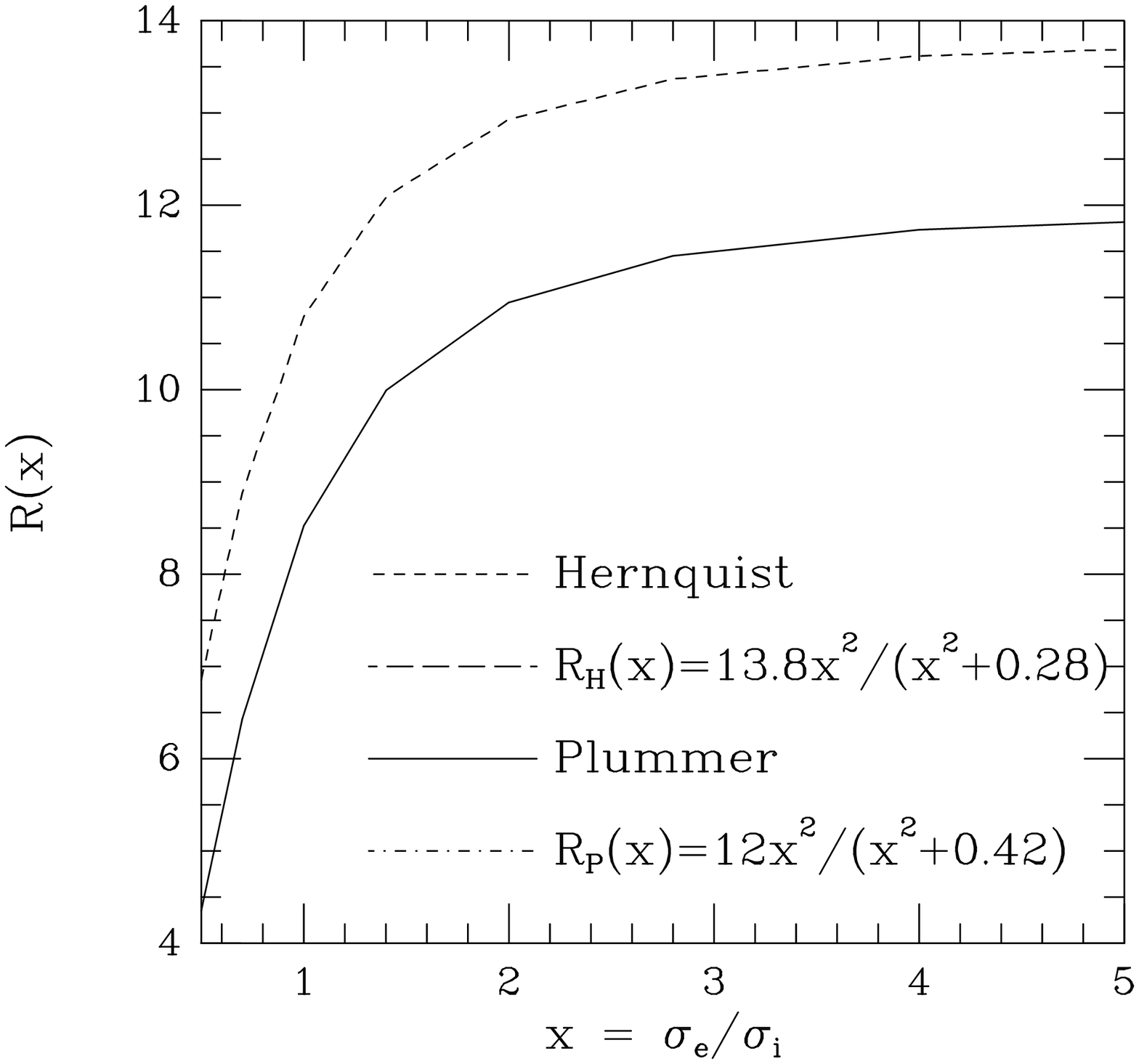]{
As Fig. 10, with the Plummer model and the Hernquist model, together
with their respective fitting formulas $R_P(x)$ and $R_H(x)$, given in 
Eqs. (34-35).
\label{fig11}}

\figcaption[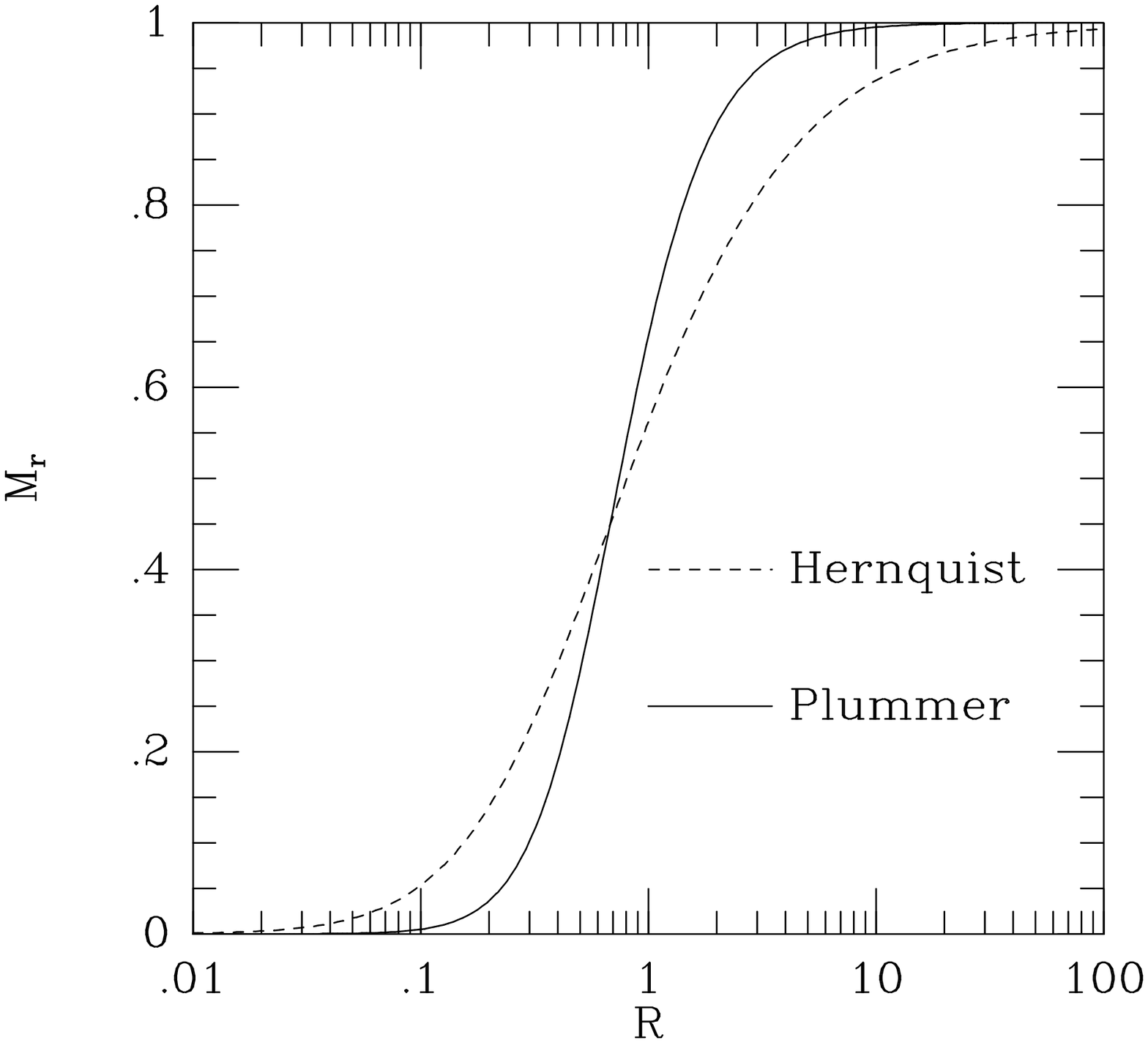]{
The cumulative mass is plotted here for the Plummer model and the
Hernquist model.  The more extended mass distribution of the latter is 
responsible for the higher merger rate of the Hernquist model in
Figs. 10 and 11.
\label{fig12}}

\figcaption[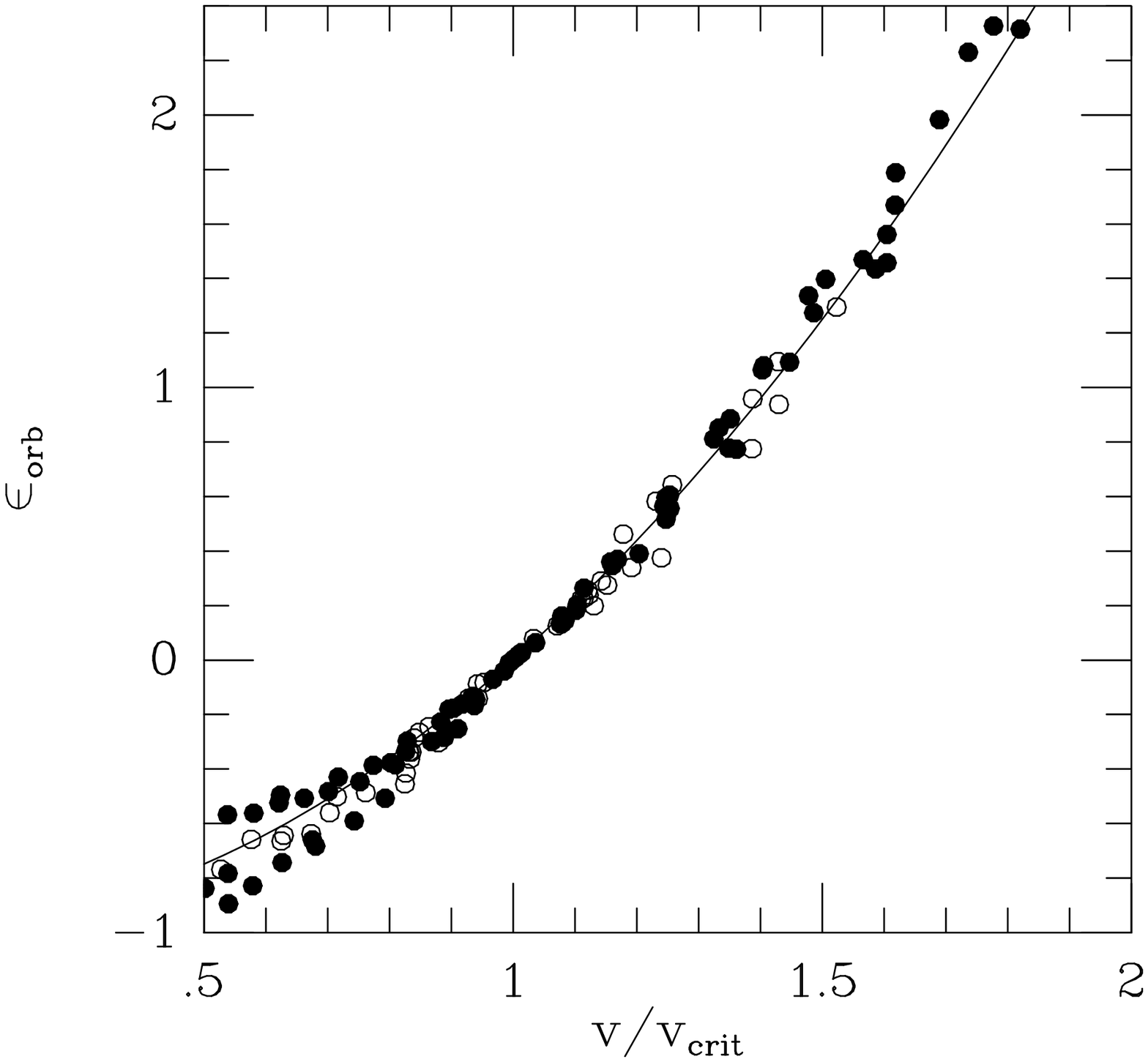]{The normalized orbital energy $\epsilon_{orb}$ (equation
(52))
after the first encounter
plotted against $v/v_{crit}$, the initial relative velocity at
infinity in units of the critical velocity for merging. Open circles
are the results for the Plummer model and filled circles are those for
the Hernquist model. The curve shows the approximation that $\Delta E$
is independent of $v$ (equation (51)).
\label{fig12b}}

\figcaption[fig13a.ps,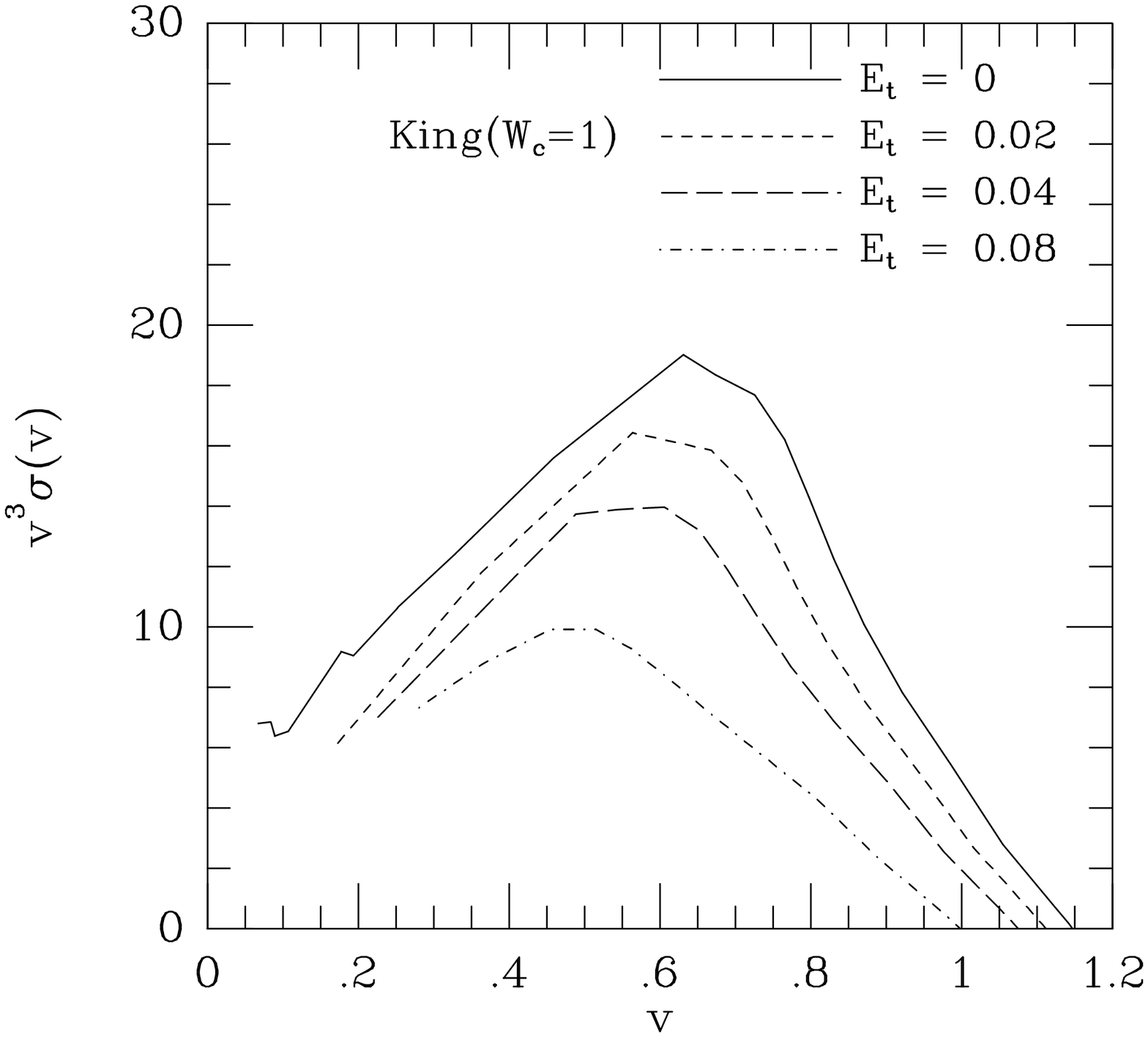,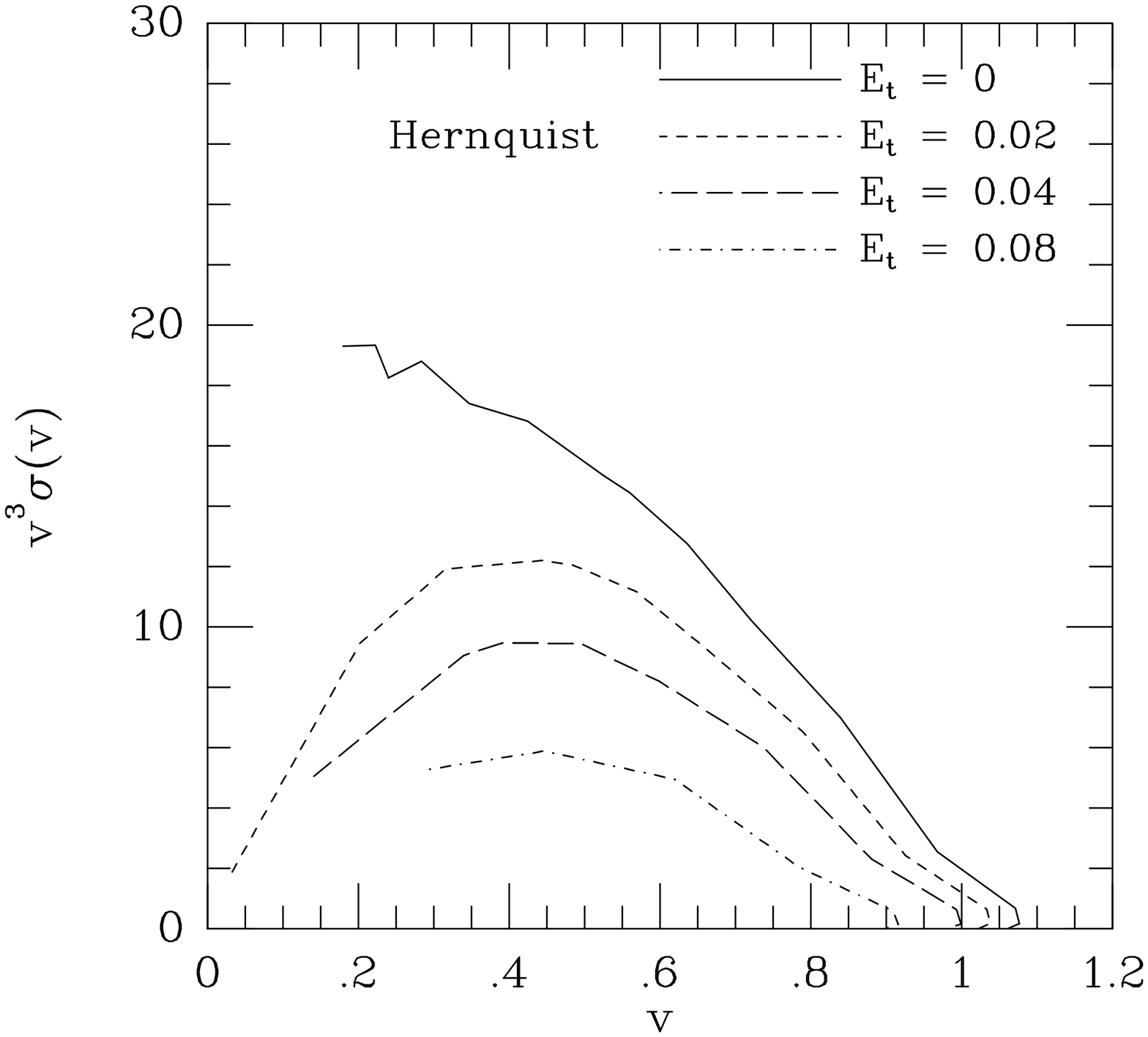]{
Same as figure 9 but for non-zero values of the critical binding
energy for tidal disruption $E_t$.  Solid, short-dashed, long-dashed,
and dot-dashed curves are results for $E_t=0$, 0.02, 0.04, 0.08, 
respectively. (a)  Plummer, (b) King model with $W_c = 1$, (c)
Hernquist model.
\label{fig13}}

\figcaption[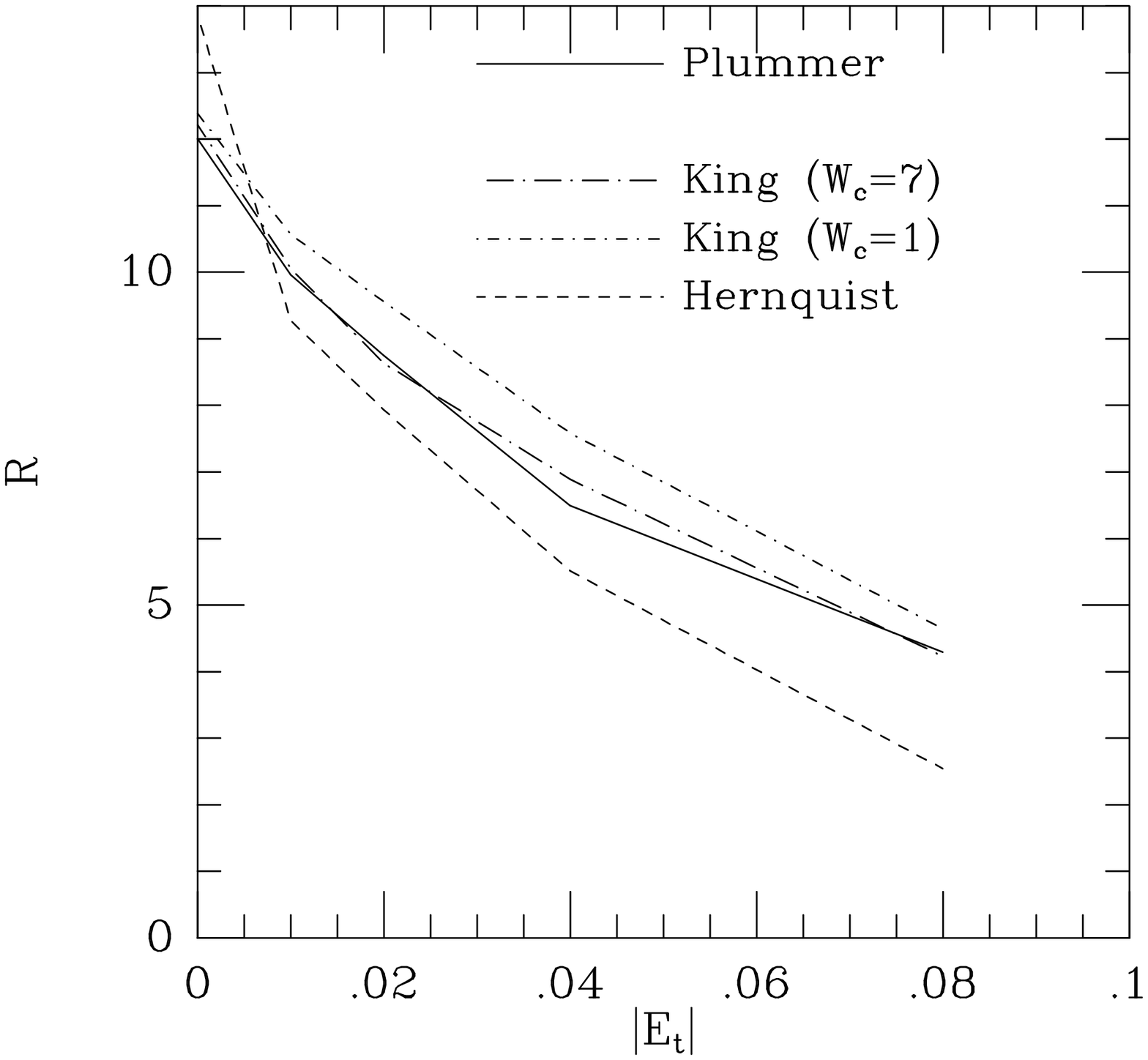]{
Merger rate $R(x)$ as a function of the critical binding
energy for tidal disruption $E_t$.  Solid, short-dot-dashed, long-dot-dashed,
and short-dashed curves are results for Plummer ($r_{cut}=23$), 
King ($W_c=1$), King ($W_c=7$),  and Hernquist models, respectively. 
\label{fig14}}

\clearpage

\plotone{fig1a.ps}

\plotone{fig1b.ps}

Figure 1

\plotone{fig2.ps}

Figure 2

\plotone{fig3.ps}

Figure 3

\plotone{fig4a.ps}

\plotone{fig4b.ps}

Figure 4

\plotone{fig5.ps}

Figure 5

\plotone{fig6.ps}

Figure 6

\plotone{fig7a.ps}

\plotone{fig7b.ps}

Figure 7

\plotone{fig8.ps}

Figure 8

\plotone{fig9.ps}

Figure 9

\plotone{fig10.ps}

Figure 10

\plotone{fig11.ps}

Figure 11

\plotone{fig12.ps}

Figure 12

\plotone{fig12b.ps}

Figure 13

\plotone{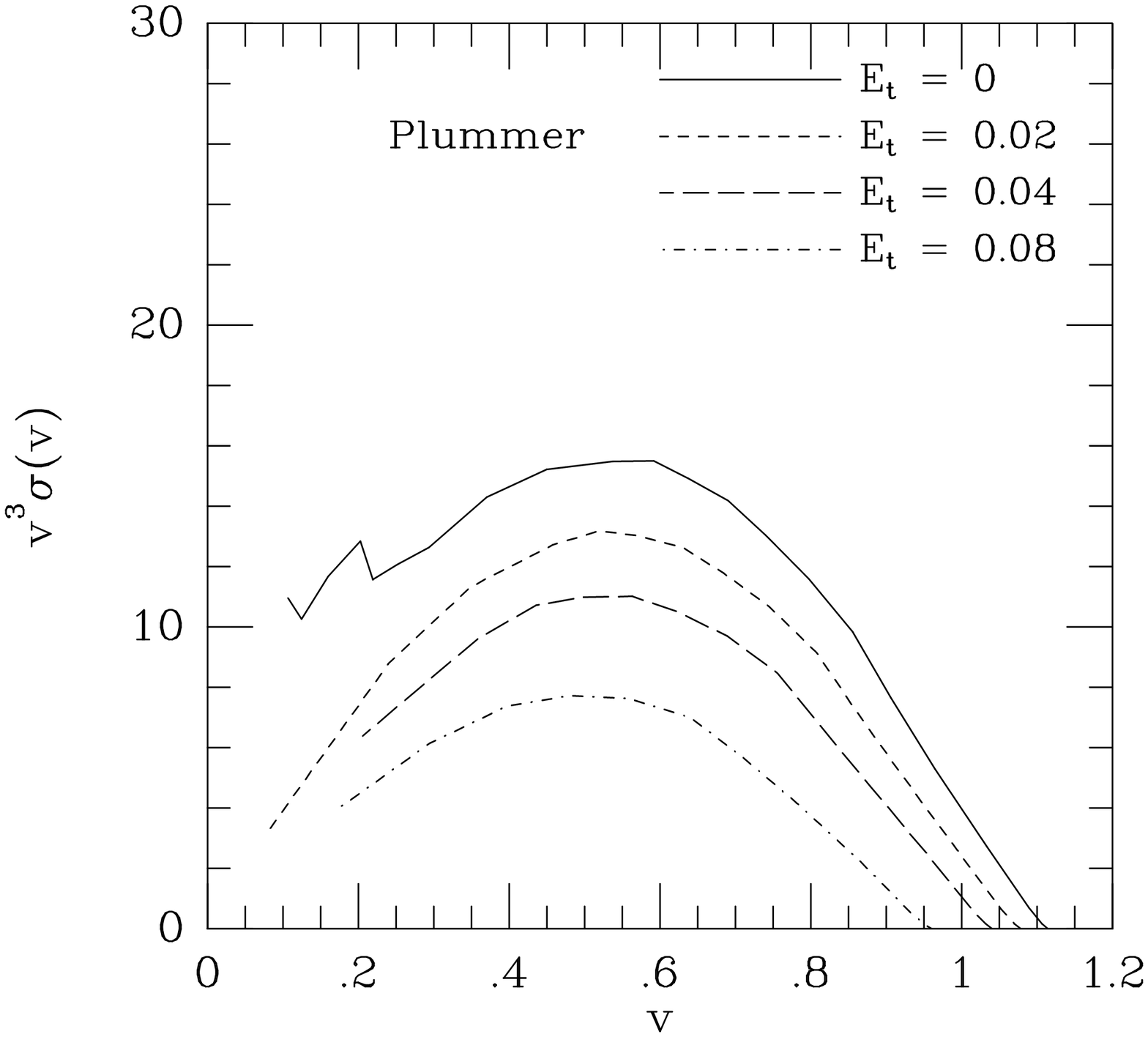}

\plotone{fig13b.ps}

\plotone{fig13c.ps}

Figure 14

\plotone{fig14.ps}

Figure 15

\end{document}